\documentclass[journal,twoside]{IEEEtran}
\usepackage{amsmath,graphicx,cite,epsfig,bm}
\usepackage{multirow}
\usepackage{booktabs}
\usepackage{float} 
\usepackage{subfigure} 
\usepackage{caption}
\usepackage{threeparttable}
\usepackage{diagbox}
\usepackage{url}

\ifCLASSINFOpdf

\else

\fi

\begin{document}
%
\title{Stereoscopic Omnidirectional Image Quality Assessment Based on Predictive Coding Theory}
%
%
%

\author{Zhibo Chen,~\IEEEmembership{Senior~Member,~IEEE}, Jiahua Xu, Chaoyi Lin and Wei Zhou  
\thanks{The authors are with the CAS Key Laboratory of Technology in Geo-Spatial Information Processing and Application System, University of Science and Technology of China, Hefei, Anhui, 230027, China (e-mail:chenzhibo@ustc.edu.cn).}
\thanks{}
\thanks{}}

%
%

\markboth{}%
{Chen \MakeLowercase{\textit{et al.}}: Stereoscopic Omnidirectional Image Quality Assessment Based on Predictive Coding Theory}
%



\maketitle

\begin{abstract}
Objective quality assessment of stereoscopic omnidirectional images is a challenging problem since it is influenced by multiple aspects such as projection deformation, field of view (FoV) range, binocular vision, visual comfort, etc. Existing studies show that classic 2D or 3D image quality assessment (IQA) metrics are not able to perform well for stereoscopic omnidirectional images. However, very few research works have focused on evaluating the perceptual visual quality of omnidirectional images, especially for stereoscopic omnidirectional images. In this paper, based on the predictive coding theory of the human vision system (HVS), we propose a stereoscopic omnidirectional image quality evaluator (SOIQE) to cope with the characteristics of 3D 360-degree images. Two modules are involved in SOIQE: predictive coding theory based binocular rivalry module and multi-view fusion module. In the binocular rivalry module, we introduce predictive coding theory to simulate the competition between high-level patterns and calculate the similarity and rivalry dominance to obtain the quality scores of viewport images. Moreover, we develop the multi-view fusion module to aggregate the quality scores of viewport images with the help of both content weight and location weight. The proposed SOIQE is a parametric model without necessary of regression learning, which ensures its interpretability and generalization performance. Experimental results on our published stereoscopic omnidirectional image quality assessment database (SOLID) demonstrate that our proposed SOIQE method outperforms state-of-the-art metrics. Furthermore, we also verify the effectiveness of each proposed module on both public stereoscopic image datasets and panoramic image datasets.
\end{abstract}

\begin{IEEEkeywords}
Image quality assessment, stereoscopic omnidirectional image, predictive coding theory, binocular rivalry, field of view, human vision system, parametric model. 
\end{IEEEkeywords}

%
\IEEEpeerreviewmaketitle

\section{Introduction}
%
%
%
%
\IEEEPARstart{W}{ith} the fast proliferation of Virtual Reality (VR) technologies, panoramic images and videos have been applied in plenty of application scenarios, such as film and television, broadcast live, cultural relic protection, product design, automatic driving, business marketing, medical examination, education, etc \cite{vrwhitepaper2017}. According to \cite{diemer2015impact}, user experience is one of the motivations for the development of VR technologies and applications. Thus, the quality assessment of VR contents has become increasingly important to maximize the perceptual experience in each stage ranging from content acquisition, format conversion, compression, transmission to display.  

Facebook released an open-source 3D-360 video capture system Facebook Surround 360 \cite{ cabral2016introducing} in 2016, which makes natural stereoscopic omnidirectional images and videos become available to consumers. Moreover, with the development of 5G, much higher bandwidth can be utilized to transmit VR contents, it is a trend that growing panoramic images and videos will be rendered with 3D format in the near future \cite{vrwhitepaper2017}. Although an in-depth study of image quality assessment (IQA) has been conducted in recent years \cite{wang2006modern}, there is still a lack of effort to predict the perceptual image quality of panoramic images, especially for stereoscopic omnidirectional images.

Similar to other image formats, the quality assessment of 3D panoramic images can be generally divided into two categories, namely subjective IQA and objective IQA \cite{seshadrinathan2010study}. Although subjective opinion provides the ultimate perceptual quality evaluation, it is limited in real applications due to the inconvenience and high cost of subjective evaluation. Therefore, it is indispensable to develop an effective objective image quality assessment algorithm which can automatically predict the perceived image quality of 3D 360-degree images. 

In order to investigate stereoscopic omnidirectional image quality assessment (SOIQA), we need to dig deeper to find out the similarities and differences among stereoscopic image quality assessment (SIQA), omnidirectional image quality assessment (OIQA) and SOIQA. SOIQA is a combination of OIQA and SIQA, it has the characteristics of both stereoscopic and omnidirectional images. As a result, SOIQA cares more about projection deformation, field of view (FoV) range, binocular perception, visual comfort, etc. Firstly, users browse panoramic images in the form of a spherical surface when wearing head-mounted display (HMD), but VR images cannot be transmitted as a sphere. Then, we need to convert the VR contents into 2D format, which is friendly to standard encoders and decoders. The most common format used in encoding is equirectangular projection (ERP) format \cite{ERP} which has a problem of pixel redundancy in polar regions. Other formats like Cubemap projection \cite{greene1986environment} would break pixel connectivity though reducing pixel redundancy. Projection deformation is introduced during the format conversion process. Secondly, in contrast to conventional 2D images, panoramic images have an unlimited field of view, users can freely change their viewing directions to explore the whole scene. However, only contents inside the viewport are visible at a time. Thirdly, binocular perception is the characteristic of 3D images as well as 3D panoramic images. Apart from depth perception, binocular fusion, rivalry or suppression might happen if there exist differences in the signals the two eyes perceived \cite{ chen2018blind }. Finally, visual discomfort is caused by the confliction between human vision and cognition \cite{ tam2011stereoscopic }. Long-term binocular rivalry or VR sickness  \cite{ kim2019vrsa } caused by fast motion will definitely cause visual discomfort. These characteristics make SOIQA an intractable challenge.

\begin{figure}[htbp]
    \centering
    \subfigure[]{
        \includegraphics[height=3.4cm]{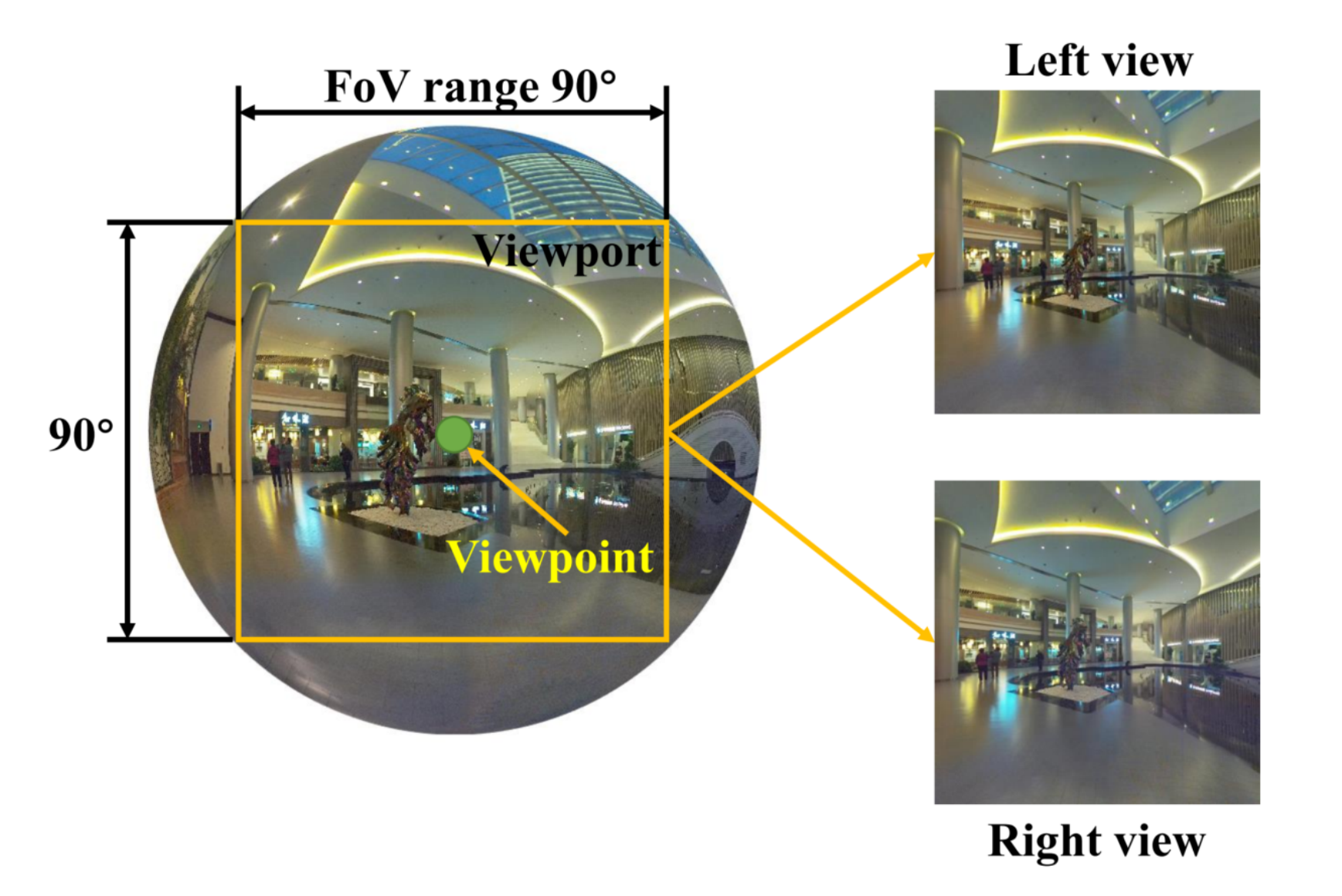}
    }
    \subfigure[]{
        \includegraphics[height=3.4cm]{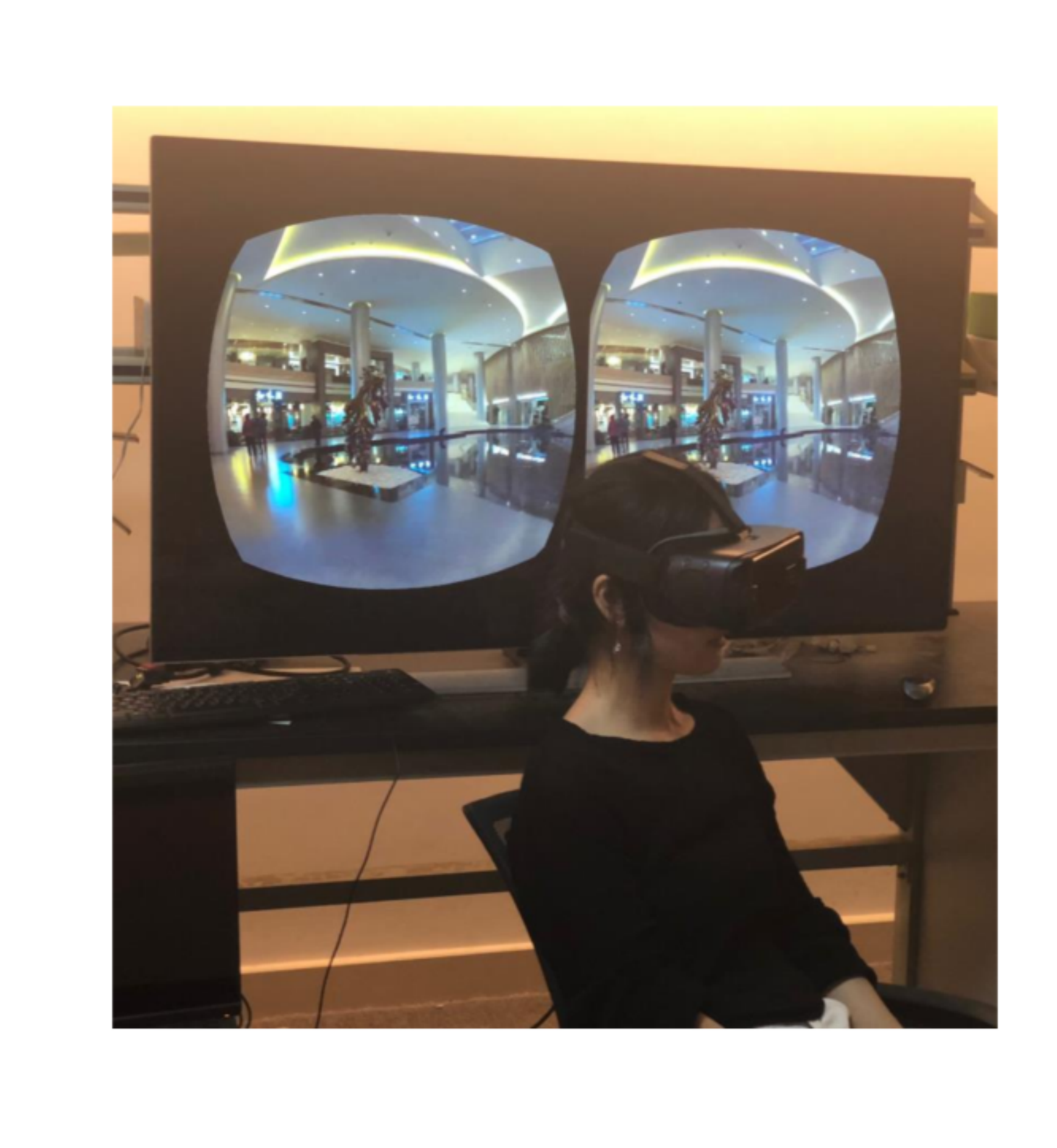}
    }
    \caption{An example of the 3D viewport image. (a) Generation of the left and right viewport images given the viewpoint and FoV range. (b) The left and right view for 3D viewport image in the realistic environment.}
    \label{s2v} 
\vspace{-0.5cm} 
\end{figure}

Based on the above characteristics of stereoscopic omnidirectional images, we convert the omnidirectional images into several viewport images given viewpoints and FoV range. An example of the 3D viewport image generation is given in Fig. \ref{s2v}(a) and Fig. \ref{s2v}(b) shows the left and right viewport images in the real environment. After conversion, the projection deformation could be alleviated owing to the reduced stretched regions and the viewport images derived from different viewpoints can still reconstruct the $360^{\circ}\times180^{\circ}$ scenery. Therefore, the problem of SOIQA is decomposed into SIQA of viewport images and quality aggregation.

Up to now, researchers have carried out studies on SIQA. In the early stage, it has been explored to combine the quality scores of two single views to predict perceived quality of 3D images based on existing 2D IQA metrics such as PSNR, SSIM \cite{ wang2004image }, MS-SSIM \cite{ wang2003multiscale } and FSIM \cite{ zhang2011fsim }. Yasakethu \emph{et al.} \cite{yasakethu2008quality} applied 2D metrics, including PSNR, SSIM and VQM \cite{ pinson2004new} to left and right view images separately and then averaged to a final score. Benoit \emph{et al.} \cite{benoit2009quality} used 2D IQA metrics to calculate the quality of left and right view images as well as disparity map. Then these quality scores were combined to estimate an overall 3D image quality. You \emph{et al.} \cite{you2010perceptual} leveraged a variety of 2D IQA metrics on stereo pairs and disparity map, then pooled them to predict quality of stereoscopic images. 

The above SIQA metrics show high performance on symmetrically distorted stereoscopic images but fail to evaluate asymmetrically distorted stereoscopic images. The asymmetrical distortion means that left and right views in stereoscopic images are impaired by different types or levels of degradation and it often leads to binocular rivalry in which perception alternates between different views \cite{levelt1965binocular}. In recent years, plenty of works have applied the binocular rivalry model to SIQA, which have achieved high performance on asymmetrically distorted stimulus and proved the effectiveness of introducing the binocular rivalry model to SIQA. 

In the literature, binocular rivalry is interpreted by a low-level competition between the input stimulus and the competition is related to the energy of stimuli \cite{ohzawa1998mechanisms, levelt1965binocular, ding2006gain}. It is believed that stimulus with higher energy will gain rivalry dominance in the competition. For example, if energy of left view image is greater than that of right view, then the left view will be dominant during the rivalry. Based on this psychophysical finding, some energy based binocular models have been proposed and applied to SIQA. Ryu \emph{et al.} \cite{ryu2014no} developed a binocular perception model considering the asymmetric properties of stereoscopic images. Chen \emph{et al.} \cite{chen2013full} proposed a full-reference metric that combined left and right views with disparity map into a cyclopean image. Afterwards, a nature scene statistics (NSS) based metric was presented by Chen \emph{et al.} \cite{chen2013no}, which was a no-reference method and support vector regression (SVR) \cite{smola2004tutorial} was used to predict the final score. These metrics need the disparity map, which is usually time-consuming and the performance is related to the stereo matching algorithm. Lin \emph{et al.} \cite{lin2014quality} incorporated binocular integration behaviour into existing 2D models for enhancing the ability to evaluate stereoscopic images. Wang \emph{et al.} \cite{wang2015quality} built a 3D IQA database and proposed a full-reference metric for asymmetrical distortion evaluation based on energy weighting.

Moreover, some existing psychophysical and neurophysiological studies have tried to explain and model the binocular rivalry phenomenon by predictive coding theory \cite{dayan1998hierarchical, hohwy2008predictive,leopold1996activity}. It is a popular theory, which is about how brain processes sensing visual stimuli. According to the predictive coding theory, the human vision system (HVS) tries to match bottom-up visual stimuli with top-down predictions \cite{spratling2017review, friston2003learning, friston2005theory}. Compared to the conventional perspective on binocular rivalry which believes competition is low-level inter-ocular competition in early visual cortex, the binocular rivalry models based on predictive coding stress more on high-level competition \cite{leopold1996activity}. Predictive coding theory has achieved some success in accounting for the response properties of the HVS \cite{rao1999predictive, spratling2012predictive, spratling2010predictive} and it fits with a wide range of neurophysiological facts \cite{shipp2013reflections,srinivasan1982predictive,kilner2007predictive,atal1979predictive,vuust2009predictive}. Therefore, we believe applying the binocular rivalry model based on predictive coding theory to SIQA is more in line with the HVS and can achieve more reliable and interpretable performance than the traditional binocular rivalry model.

As previously stated, SOIQA is a combination of SIQA and OIQA. SIQA is handled with predictive coding based binocular rivalry model in our method. Another significant aspect of SOIQA is how to predict the quality of a 360-degree image. Quality assessment of panoramic contents has attracted extensive attention recently due to the rapid development of VR technologies. To further investigate OIQA, several databases \cite{sun2018large,duan2018perceptual} consisting of various distortions were built for the design of objective metrics to automatically predict image quality. Firstly, some PSNR based metrics were proposed to evaluate the quality of panoramic images, namely spherical PSNR (S-PSNR), weighted-to-spherically-uniform PSNR (WS-PSNR), craster parabolic projection PSNR (CPP-PSNR). Instead of calculating PSNR directly on projected images, Yu \emph{et al.} \cite{yu2015framework} proposed S-PSNR to overcome the oversampling drawback of redundant pixels and selected uniformly distributed points on the sphere. However, these sampled points are usually fixed and less than spherical pixels which may cause information loss. Then, Sun \emph{et al.} \cite{sun2017weighted} developed WS-PSNR that can be directly calculated on 2D format without converting to other formats. The original error map was multiplied by a weight map that can reduce the influence of stretched areas, but WS-PSNR cannot work across different formats. Moreover, CPP-PSNR \cite{ zakharchenko2016quality} was put forward to resolve the problems of S-PSNR and WS-PSNR. It utilized all the pixels on the sphere surface and can be applied to different projection formats, but interpolation was introduced during format conversion which would lower the precision. Though these PSNR based objective metrics can be easily integrated into codecs, their relevance with subjective perception is still quite low.

Therefore, some perception-driven quality assessment models for VR contents were designed via machine learning scheme. Kim \emph{et al.} \cite{kim2019deep} presented a deep network consisting of VR quality score predictor and human perception guider. The VR quality score predictor was trained to accurately assess the quality of omnidirectional images and fool the guider while the human perception guider aimed to differentiate the predicted scores and subjective scores. Yang \emph{et al.} \cite{yang20183d} applied 3D convolutional neural networks (3D CNN) \cite{ji20133d} to blindly predict 3D panoramic video quality. It took the difference frames between left and right view images as inputs which can reflect distortion and depth information. However, the above models were trained on 2D patches from ERP format that conflicted with the actual viewing experience. As inspired by \cite{xu2018assessing,xu2018predicting}, viewport images play a significant role in OIQA, we therefore build our model based on viewport images. 

In addition, Yang \emph{et al.} \cite{yang2017objective} leveraged multi-level quality factors with region of interest (ROI) analysis to estimate the quality of panoramic videos using back propagation (BP) neural network. It inspires us that different viewport images share various weights in one omnidirectional image. According to \cite{rai2017saliency,rai2017dataset}, head and eye tracking data could be utilized to acquire saliency information that is beneficial to OIQA. Consequently, location weight and content weight are introduced for aggerating image quality scores of separate viewport images.

In this paper, we propose a biologically plausible binocular rivalry module based on predictive coding theory for assessing the perceptual quality of stereoscopic viewpoint images. To the best of our knowledge, it is the very first work to introduce predictive coding theory into modeling binocular rivalry in SIQA as well as SOIQA. Specifically, binocular rivalry is simulated as the competition between high-level patterns rather than low-level competitions since the principle of the HVS process is to match bottom-up visual stimuli with top-down predictions \cite{leopold1996activity}. Moreover, a multi-view fusion module is developed to integrate quality scores of viewport images through both location weight and content weight scheme. The binocular rivalry module and the multi-view fusion module can be applied to 3D images and 2D panoramic images respectively. Finally, the two modules form the stereoscopic omnidirectional image quality evaluator (SOIQE) which can accurately predict the visual quality of stereo 360-degree images. It is a parametric model without necessary of regressive learning and each parameter in this model corresponds to a clear physical meaning. We test SOIQE on the self-built public stereoscopic omnidirectional image quality assessment database (SOLID) \cite{xu2018subjective} and the experimental results show its high correlation with human judgements. Besides, due to the lack of other 3D panoramic image databases, the generalization and robustness of our method are verified by the performance evaluation on two well-known 3D image databases and two public 2D omnidirectional image databases. The SOLID database and the source code of SOIQE are available online for public research usage $\footnote{\url{http://staff.ustc.edu.cn/~chenzhibo/resources.html}}$.

The rest of this paper is organized as follows. The predictive coding theory is reviewed in Section II. Section III introduces the proposed stereoscopic omnidirectional image quality evaluator for SOIQA in details. We present the experimental results and analysis in section IV and conclude the paper in Section V.

\begin{figure*}[thbp]
  \vspace{-0.5cm} 
  \centerline{\includegraphics[width=15cm]{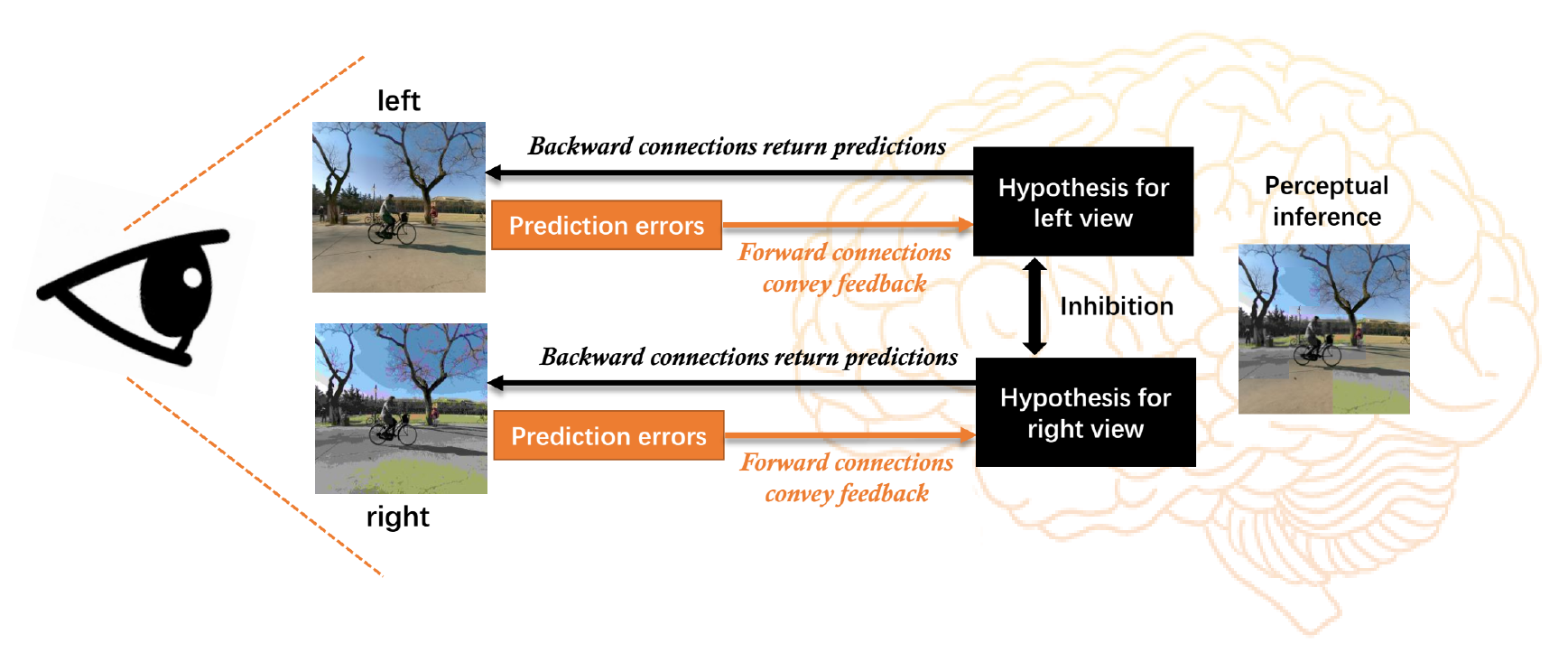}}
  \vspace{-0.3cm} 
  \caption{Simplified schematic of predictive coding theory on binocular rivalry. The black arrow is the top-down prediction from higher level and the orange arrow is the bottom-up error signals. The Hypothesis for left and right view will compete with each other and the brain will obtain the final perceptual inference.}
  \centering
\label{pc_ex} 
\end{figure*}

\section{Predictive Coding Theory}

Representing the environmental causes of its sensory input is the core task for the brain \cite{hohwy2008predictive}. Given a sensory input, the neural computation system will predict what the cause of sensory input is. Then, the perceptual content is determined by the hypothesis (\emph{i.e.} predicted cause) which generates the best prediction. However, it is computationally difficult because the hypothesis is difficult to predict. The hierarchical Bayesian inference using the generative model can deal with these challenges by furnishing formal constraints on the mapping between hypothesis and effect \cite{hohwy2008predictive}. In this section, we will introduce a predictive coding model and propose a binocular rivalry model based on the predictive coding theory.

\subsection{Predictive Coding Model}
The simplest predictive coding model is the linear predictive coding (LPC) in digital signal processing \cite{vaseghi2008advanced}. It is first applied to explain efficient encoding in the retina, and then subsequently used to model the approximate Bayesian inference in the HVS.

Rao and Ballard \cite{rao1999predictive} proposed a hieratical model that the feedback from higher level carries the predictions of lower level stimuli while the feedforward carries the residual errors between the stimuli and the predictions. In this paper, we adopt Rao's model because of its powerful ability for representing natural images. Note that the predictive coding model introduced here is for monocular vision and it is the basis of our binocular rivalry model. Given an image $\bm{I}$, it is assumed that the cortex tries to represent the image in terms of hypothesis, which is represented by a vector $\bm{r}$. This relationship can be modeled by a generative model that the image $\bm{I}$ is generated by a combination of the basis vectors:
\begin{equation}\label{2.1}
\bm{I}=f\left ( \bm{Ur} \right )+\bm{n},
\end{equation}
where $f\left (\cdot \right )$ is the activation function, $\bm{n}$ is stochastic noise and $\bm{U}$ is a dictionary. In this model, $f\left ( \bm{Ur} \right)=f\left ( \sum_{j=1}^{k}\bm{U}_{j}r_{j} \right ) $, the coefficients $r_{j}$ correspond to the firing rates of neurons and the basis vectors $\bm{U}_{j}$ correspond to the synaptic weights of neurons. The basis vectors in the dictionary $\bm{U}$ are also called patterns in the predictive coding model. Binocular rivalry model based on predictive coding theory stresses more on the high-level competition between the patterns rather than low-level signals \cite{leopold1996activity}. Thus, the patterns are important in our binocular rivalry model.

Given an image $\bm{I}$, in order to estimate the coefficient $\bm{r}$ and the dictionary $\bm{U}$, the optimization function is as follows \cite{rao1999predictive}:
\begin{equation}\label{2.2}
E=\frac{1}{\sigma ^{2}}\left ( \bm{I}-f\left ( \bm{Ur} \right ) \right )^{T}\left ( \bm{I}-f\left ( \bm{Ur} \right ) \right )+g\left ( \bm{r} \right )+h\left ( \bm{U} \right ),
\end{equation}
where $T$ represents the transpose of a vector, $ g\left ( \bm{r} \right )=\alpha \sum _{i}log\left ( 1+r_{i}^{2} \right )$ and $ h\left ( \bm{U} \right )=\lambda  \sum _{i,j}U_{i,j}^{2}$ are the regularization terms for $\bm{r}$ and $\bm{U}$, respectively. Here, the noise $\bm{n}$ is assumed as Gaussian with zero mean and variance $\sigma^{2}$. Then, the optimal $\bm{U}$ and $\bm{r}$ are obtained using gradient descent algorithm. As a result, the response $\bm{r}$ indicates the current estimation of the input image $\bm{I}$ and $\bm{U}^{T}$ represents the synaptic weights. The feedback neurons convey the prediction $ f\left ( \bm{Ur} \right )$ to the low level, and then the difference $ \bm{I}-f\left ( \bm{Ur} \right )$ between the current stimuli $\bm{I}$ and the top-down prediction $\bm{r}$ is calculated by the error-detecting neurons.

\subsection{Binocular Rivalry Model Based on Predictive Coding}

The binocular rivalry is a phenomenon in which perception alternates between left and right views. This phenomenon is highly related to SIQA because of the possible asymmetrical distortion in stereo images. Therefore, it is important to apply the binocular rivalry model to SIQA. There have emerged some research works about using predictive coding theory to model binocular rivalry phenomenon in recent years. A theoretical framework for the computational mechanism of binocular rivalry was proposed in an epistemological review \cite{hohwy2008predictive}. Compared to the conventional perspective on binocular rivalry which believes competition is low-level inter-ocular competition in early visual cortex, the binocular rivalry stresses more on high-level competition \cite{leopold1996activity}. In \cite{dayan1998hierarchical}, a hierarchical model of binocular rivalry was proposed based on the hypothesis that it is competition between top-down predictions for input stimuli rather than direct competition of stimuli. However, most existed  binocular rivalry models are not developed for natural scene image. Therefore, they cannot be applied to image quality assessment directly, such as the model proposed in \cite{leopold1996activity}.

In this paper, we develop a binocular rivalry model based on predictive coding theory for SIQA according to a general theoretical framework introduced in \cite{hohwy2008predictive}. From the Bayesian perspective, human will perceive the content because the corresponding hypothesis has the highest posterior probability \cite{friston2002functional,kersten2004object}. In Fig. \ref{pc_ex}, given a stereoscopic image, our brain will first determine a hypothesis that can best predict the corresponding stimulus, which is regarded as the likelihood. Besides, the perceptual inference also depends on the prior probability of hypotheses, which is about how probable the hypothesis is. Then, the hypothesis for left view and right view will compete with each other. The hypothesis which has the highest posterior probability will win and the corresponding stimulus will be dominant during the rivalry while the hypothesis with lower posterior probability will be inhibited \cite{hohwy2008predictive}.

\section{Proposed Stereoscopic Omnidirectional Image Quality Evaluator (SOIQE)}

\begin{figure*}[htbp]
  \centerline{\includegraphics[width=18cm]{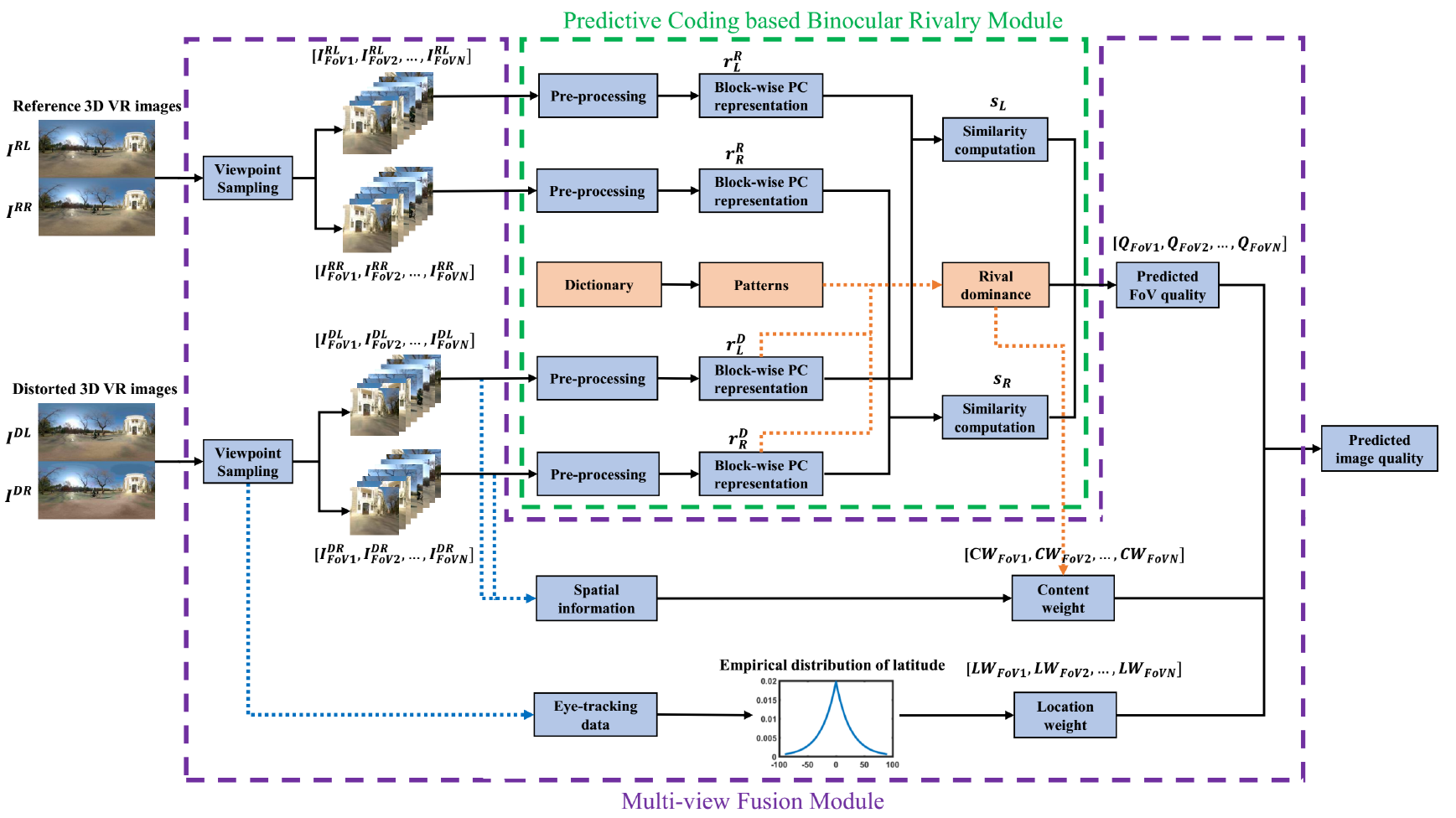}}
  \caption{The architecture of our proposed stereoscopic omnidirectional image quality evaluator (SOIQE). It takes distorted and reference 3D 360-degree images as input and converts the images into $N$ stereo viewport images. In the predictive coding based binocular rivalry module (PC-BRM), quality of $N$ viewport images is computed. In the multi-view fusion module (MvFM), the final quality is calculated by weighing every viewport's quality with its location and content.}
  \centering
\label{fig:fig1}
\end{figure*}

Generally, viewing a stereoscopic omnidirectional image is actually browsing several stereo images continuously. Therefore, the problem of stereoscopic omnidirectional image quality assessment can be converted into multi-view 3D IQA inside FoV, which is the basic idea of our model. In this section, we will introduce our model in details.

\subsection{Architecture}
The framework of our proposed SOIQE is shown in Fig. \ref{fig:fig1}. It contains the predictive coding based binocular rivalry module (PC-BRM) and the multi-view module (MvFM). Given a stereo panoramic image pair in the ERP format, we first perform automatic downsampling according to \cite{wang2004image}. Then, multi-viewport images could be acquired from reprojection. PC-BRM aims to extract the high-level cause of the viewport stereo image pairs for calculation of similarity and rivalry dominance. Both hypothesis likelihood and prior are considered in the rivalry advantage. Further, the quality scores of $N$ viewport images are estimated and need to be aggregated, which can be implemented by MvFM. The content weight for each viewport image is reflected by its spatial information (SI). In addition, the centre latitude of each viewport image is utilized to calculate its corresponding location weight. Finally, the quality scores of all viewport images are fused together with the normalized content and location weight to predict stereoscopic omnidirectional image quality.

\begin{figure}[htbp]
    \centering
    \subfigure[]{
        \includegraphics[width=2.7cm]{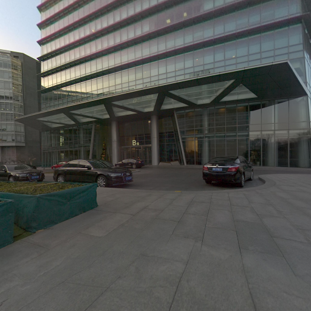}
    }
    \subfigure[]{
        \includegraphics[width=2.7cm]{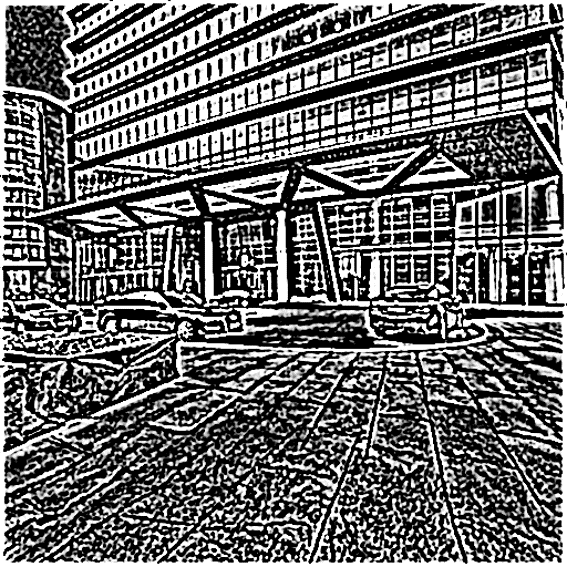}
    }
    \subfigure[]{
        \includegraphics[width=2.7cm]{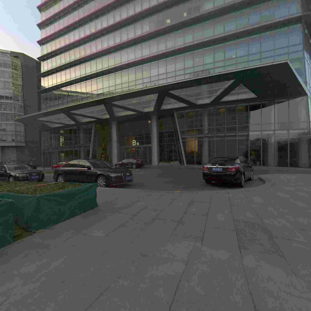}
    }
    \subfigure[]{
        \includegraphics[width=2.7cm]{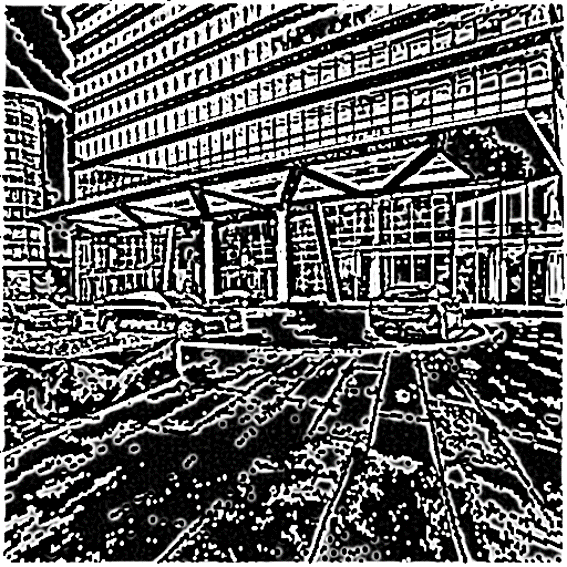}
    }  
    \caption{(a) The reference viewpoint image, (b) The preprocessed reference viewpoint image, (c) The distorted viewpoint image, (d) The preprocessed distorted viewpoint image.}
    \label{fig:fig2} 
\vspace{-0.5cm} 
\end{figure}

\begin{figure*}[htbp]
\vspace{-0.5cm} 
  \centerline{\includegraphics[width=14cm]{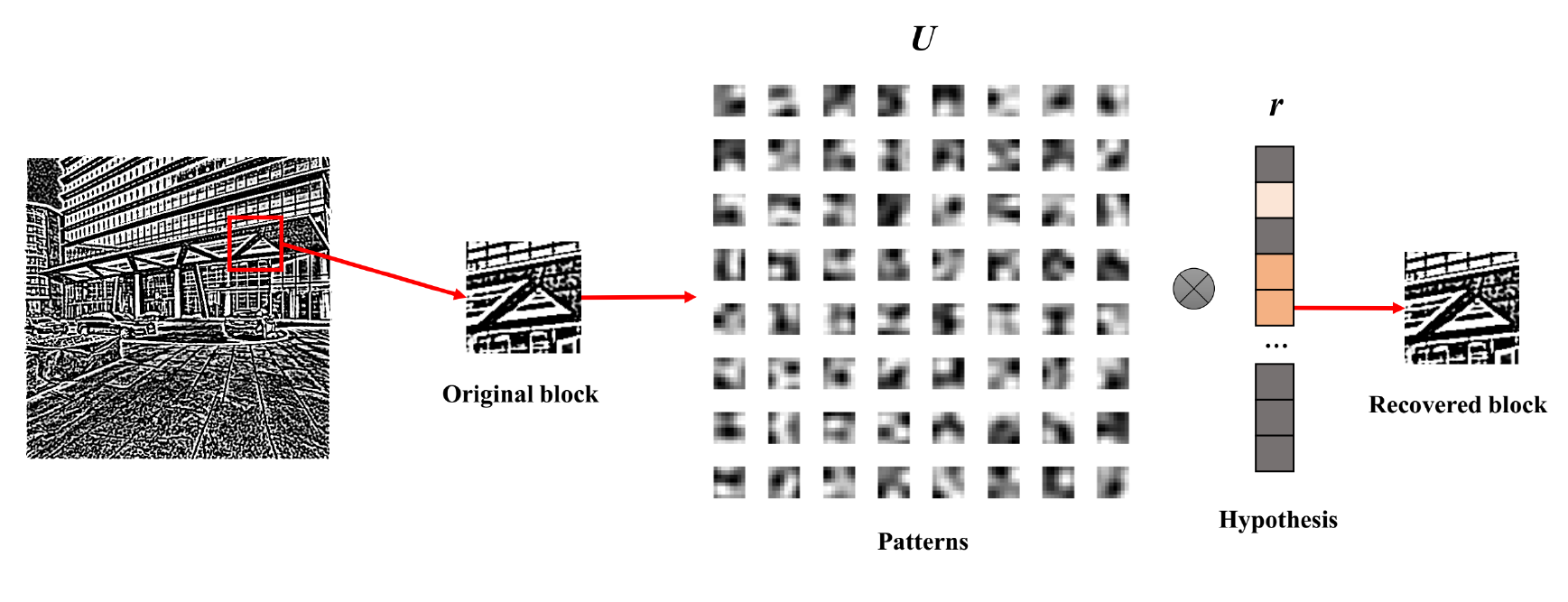}}
  \caption{Illustration of predictive coding representation. Combination of patterns and coding coefficients (hypothesis) is used to reconstruct the image block.}
  \centering
\label{fig:fig3}
\end{figure*}

\subsection{Predictive Coding Based Binocular Rivalry Module}
We use the predictive coding based binocular rivalry module to predict the quality of stereo viewport image pairs. 3D 360-degree images are converted into viewport images in order to simulate the competition between high-level patterns in binocular rivalry when viewing stereoscopic images. 
\subsubsection{\textbf{Preprocessing of Viewport Images}}
The preprocessing stage is inspired by the lateral geniculate nucleus (LGN) in the HVS. Given a viewport image $\bm{I}$, it is convolved with a Laplacian-of-Gaussian (LoG) filter $\bm{L}$, which standard deviation is equal to 1.5. The LoG filter is virtually identical to the Difference-of-Gaussian (DoG) filter, which has been traditionally used to model circular Receptive Fields in LGN \cite{spratling2010predictive}. The output of this filter is subject to a saturating non-linearity function:
\begin{equation}\label{1}
\bm{I}_{p}=tanh\left \{ 2\pi \left ( \bm{I}\otimes \bm{L} \right ) \right \}.
\end{equation}
According to this equation, the preprocessed reference and distorted stereoscopic viewpoint images are denoted as $\bm{I}_{p}^{Ref}$ and $\bm{I}_{p}^{Dis}$, respectively. The reference and distorted viewport images are depicted in Fig. \ref{fig:fig2}(a) and Fig. \ref{fig:fig2}(c). Fig. \ref{fig:fig2}(b) and  Fig. \ref{fig:fig2}(d) are the corresponding preprocessed images. Observed from Fig. \ref{fig:fig2}(d), some detailed information is lost while some additional information appears compared to Fig. \ref{fig:fig2}(b).

\subsubsection{\textbf{Predictive Coding Representation}}
We employ the predictive coding (PC) representation to extract the high-level cause of input stereo images for computing similarity and rivalry dominance. Given a stimulus, the hypothesis is predicted in this procedure. We use Rao's hierarchical model \cite{rao1999predictive} in this stage and only one level is adopted in our model with lower computation complexity.

Firstly, the dictionary $\bm{U}$ is trained by gradient descent algorithm on a 2D image quality assessment database \cite{sheikh2006statistical}. Since the panoramic image has been converted into viewport images and predictive coding model is used to calculate the hypothesis for each view, inputs of this model are 2D viewport images which are similar to conventional 2D natural images. Hence, we can leverage the 2D image database to train the dictionary and avoid overfitting issue if training is taken on tested images. The influence of dictionary size will be discussed in Section IV.

Based on the dictionary, the predictive coding model is used to process images. Given a pre-processed image pair $\bm{I}_{p}^{Ref}$ and $\bm{I}_{p}^{Dis}$ with size $M\times N$, we partition them into non-overlapping patches with the same size $L\times L$. The blocks of preprocessed reference and distorted images are denoted by $\bm{I}_{i}^{Ref}$ and $\bm{I}_{i}^{Dis}$ respectively, where $i=1,2,\cdots ,\left\lfloor \frac{M\times N}{L\times L} \right\rfloor$, $\left\lfloor \cdot  \right\rfloor$ is the floor operation. After predictive coding representation, coding coefficients $\bm{r}_{i}^{Ref}$ and $\bm{r}_{i}^{Dis}$ for the $i$-th block in the reference image and distorted image are achieved. Fig. \ref{fig:fig3} illustrates the detailed process of predictive coding representation. As shown in this figure, the block of the original image is reconstructed by a combination of the basis vectors and coding coefficients. Corresponding to the predictive coding theory, the hypothesis is represented as the coding coefficients $\bm{r}$ and the high-level patterns are represented as basis vectors.

\subsubsection{\textbf{Similarity Calculation}}
After the predictive coding representation, similarity calculation is performed. We aim to calculate the similarity between the distorted image block $\bm{I}_{i}^{Dis}$ and reference image block  $\bm{I}_{i}^{Ref}$ in this step. For each block, the similarity $s_{i}$ is calculated as follows:
\begin{equation}\label{2}
s_{i}=\frac{1}{N}\sum_{j}\left ( \frac{2r_{ij}^{Ref}r_{ij}^{Dis}+C}{{r_{ij}^{Ref}}^{2}+{r_{ij}^{Dis}}^{2}+C} \right ),
\end{equation}
where $C$ is a constant to prevent dividing by 0, $N$ is the number of basis vectors in the dictionary, $r_{ij}^{Ref}$ represents the $j$-th elements in vector $\bm{r}_{i}^{Ref}$ and $\bm{r}_{i}^{Ref}$ is the hypothesis of the $i$-th block in reference image. $r_{ij}^{Dis}$ is similar to $r_{ij}^{Ref}$ but it represents the coding coefficient of the distorted image. Until now, we can calculate $s_{i}^{L}$ as the similarity of $i$-th block between the reference and distorted viewpoint image of left view and $s_{i}^{R}$ as that of right view.

\subsubsection{\textbf{Rivalry Dominance Allocation}}
Binocular rivalry occurs when viewing asymmetrically distorted images without reference, thus only distorted left and right view images are used to calculate the rivalry dominance. In conventional binocular rivalry model, the dominance is usually calculated by the energy of left and right views. For the binocular rivalry based on predictive coding, as we analyzed before, the hypothesis with the highest posterior probability will determine the perceptual content. Thus, the prior and likelihood of the hypothesis are important to perform perceptual inference and we utilize both of them to produce the dominance. Note that we are not going to calculate a probability, we just try to model the likelihood and prior with a quantity that has similar physical meaning.

In this module, the patterns (\emph{i.e.} basis vectors) in predictive coding are used to model the prior. The prior is about how probable the hypothesis is and it is concerned with the knowledge learned before. Considering the process of training dictionary, the trained patterns can reflect the texture or statistical property of images from the training dataset. Given a test image, if it is similar to the image in the training dataset, it can be usually reconstructed well in the predictive coding model. As the prior is related to the patterns, we use the patterns to model the prior. If the input is an image with complex texture, then the patterns used to reconstruct the input should also contain enough information to recover it. The texture complexity of the pattern can be reflected with its variance. Larger variance in the pattern means it may contain high-frequency information. The variance of each basis vector can be denoted as $Var\left ( \bm{U}_{j} \right)$, $j = 1, 2,\cdots ,N$ and the prior of hypothesis for each patch in binocular rivalry is calculated as $v_{i}$:
\begin{equation}\label{3}
v_{i}^{L}=\sum _{j}Var\left ( \bm{U}_{j} \right )r_{ij}^{L},
\end{equation}
\begin{equation}\label{4}
v_{i}^{R}=\sum _{j}Var\left ( \bm{U}_{j} \right )r_{ij}^{R},
\end{equation}
where $j$ means the $j$-th basis vector used to predict (\emph{i.e.} reconstruct) the input stimuli and $r_{ij}$ represents how much the basis vector contributes when reconstructing the input stimuli. And $r_{ij}$ is obtained from the distorted image blocks in Eq. \ref{3} and \ref{4}.

The likelihood is about how well the hypothesis predicts the input and the prediction error is used to model the likelihood in our binocular rivalry model. The prediction error reflects the difference between the stimulus and the top-down prediction. If one hypothesis can explain the stimulus primely, then it should have a smaller prediction error. That is, one hypothesis with the large likelihood means it has a small prediction error. In our algorithm, the likelihood is represented as the sum of squared error map for each block:
\begin{equation}\label{5}
EW_{i}^{L}=1-\frac{\sum{\bm{E}_{i}^{L}}^{2}}{\sum{\bm{E}_{i}^{L}}^{2}+\sum{\bm{E}_{i}^{R}}^{2}},
\end{equation}
\begin{equation}\label{6}
EW_{i}^{R}=1-\frac{\sum{\bm{E}_{i}^{R}}^{2}}{\sum{\bm{E}_{i}^{L}}^{2}+\sum{\bm{E}_{i}^{R}}^{2}},
\end{equation}
where $\bm{E}_{i}^{L}$ represents the error map of left view for the $i$-th block and $\bm{E}_{i}^{R}$ represents that of right view. $\sum\bm{E}^{2}$ denotes the sum of squared error map. The $EW_{i}^{L}$ and $EW_{i}^{R}$ are normalized between each other to avoid the effect of magnitude issue.

\subsubsection{\textbf{Viewport Image Quality Estimation}}

The prediction error can also reflect distortion characteristics. This is inspired by \cite{zhu2016stereoscopic} that if one image is distorted by Gaussian white noise, then the prediction error will also contain white noise. The distortion characteristic can be described by the variance of the error map.
\begin{equation}\label{7}
R_{i}=var\left ( \bm{E}_{i}^{2} \right ).
\end{equation}

Given the prior and likelihood, it is natural to combine them by multiplication. Thus, the final estimated quality of the viewpoint stereo image is calculated as follows:
\begin{equation}\label{8}
Q_{FoVn}=\sum _{i}\left ( v_{i}^{L} EW_{i}^{L} R_{i}^{L} s_{i}^{L} + v_{i}^{R} EW_{i}^{R} R_{i}^{R} s_{i}^{R} \right ),
\end{equation}
where $Q_{FoVn}$ is the quality of $n$-th viewport image. In order to avoid the effect of magnitude issue, the $v_{i}$ and $R_{i}$ are normalized between left and right views similar to Eq. \ref{5} and \ref{6}, respectively. In Eq. \ref{8}, it can be seen that each parameter has a clear physical meaning and these parameters are flexible to be combined for predicting the perceived quality of stereoscopic images.

\subsection{Multi-view Fusion Module}
We propose the multi-view fusion module to fuse the quality of sampled viewport images. First of all, we sample the viewpoints according to the designed strategy described in the next paragraph. Then, content weight and location weight are introduced to implement the quality fusion and calculate the final quality of a stereoscopic omnidirectional image.

\begin{figure}[htbp]
  \centerline{\includegraphics[width=8cm]{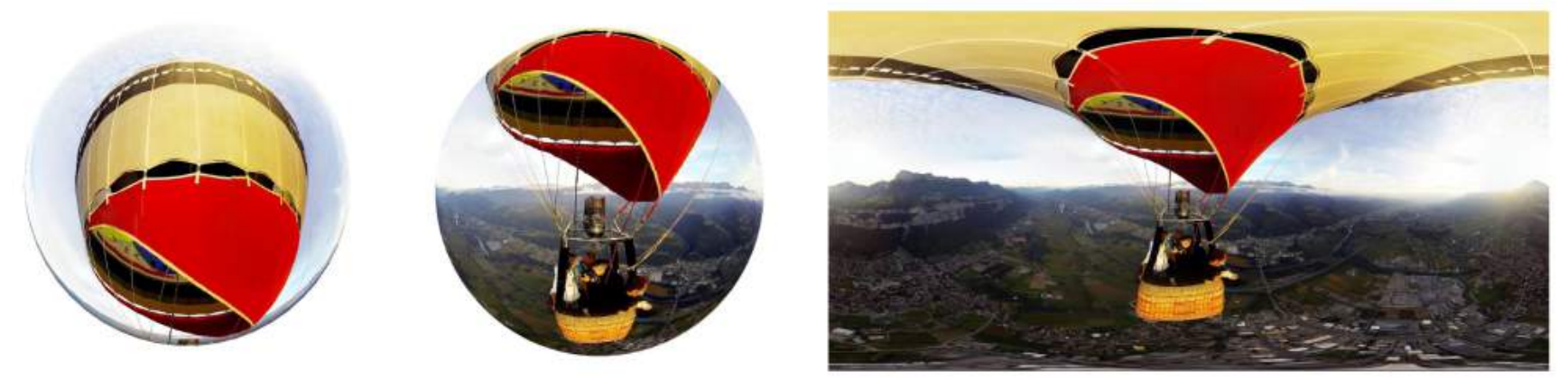}}
  \caption{Stretched polar regions in the panoramic image \cite{abbas2016gopro}.}
  \centering
\label{fig:fig4}
\end{figure}

\subsubsection{\textbf{Viewpoint Sampling Strategy}}
360-degree images are usually transmitted in ERP format which will stretch polar regions as shown in Fig. \ref{fig:fig4}. However, a panoramic image is projected onto a sphere surface when being viewed in the HMD and it differs from being projected into the ERP format. Considering the polar regions are stretched, we take a different viewpoint sampling strategy instead of uniform sampling. Firstly, $N_{0}$ viewpoints are equidistantly sampled on the equator and the angle between two adjacent viewpoints is computed as $\theta =\frac{360}{N_{0}}$. Then, $N_{1}$ viewpoints of $\theta$ degrees north (south) latitude could be sampled uniformly as follows:
\begin{equation}\label{9}
N_{1}=\left \lfloor N_{0} cos\theta \right \rfloor ,
\end{equation}

\begin{figure}[htbp]
    \centering
    \subfigure[]{
        \includegraphics[height=2.5cm]{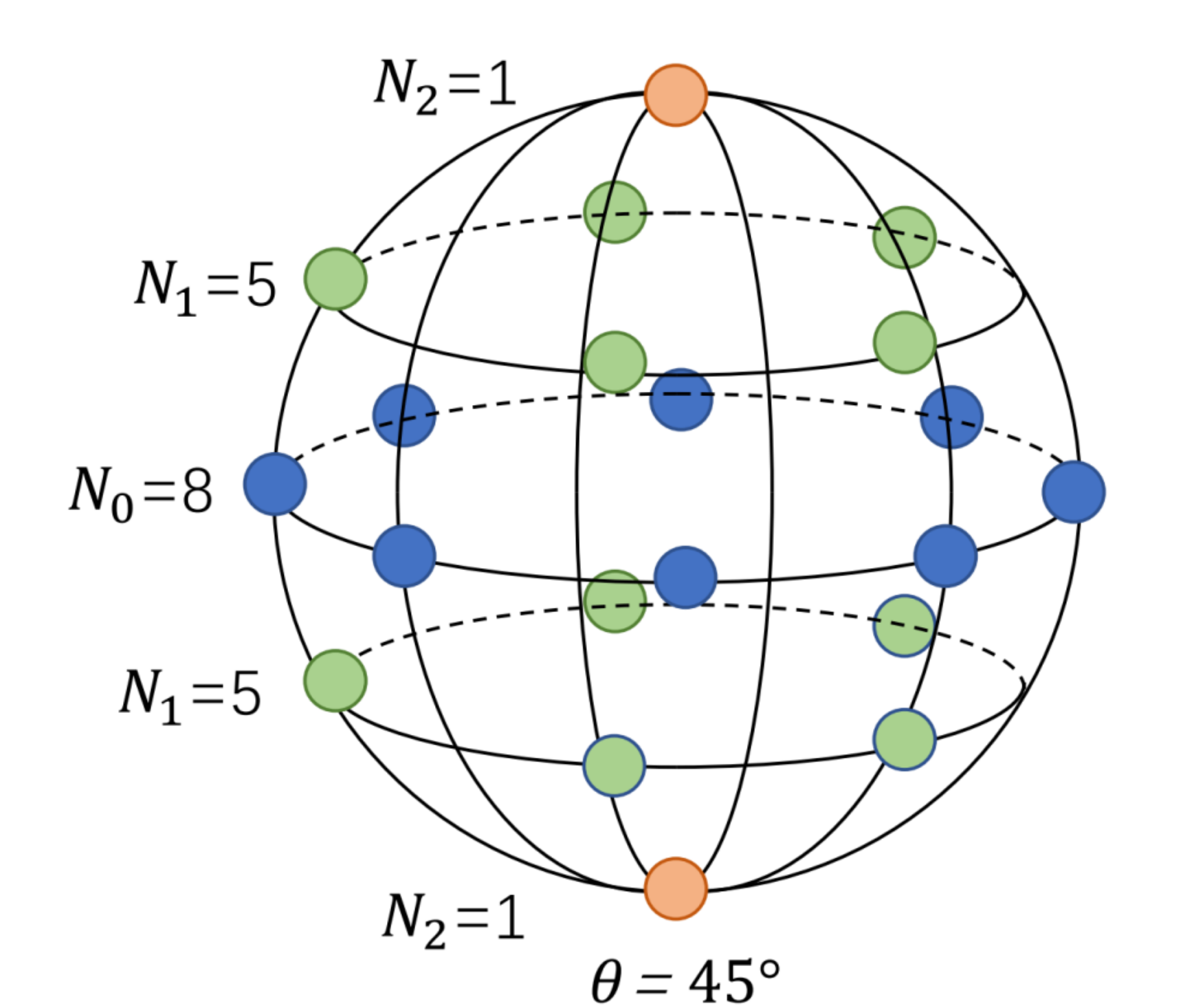}
    }
    \subfigure[]{
        \includegraphics[height=2.5cm]{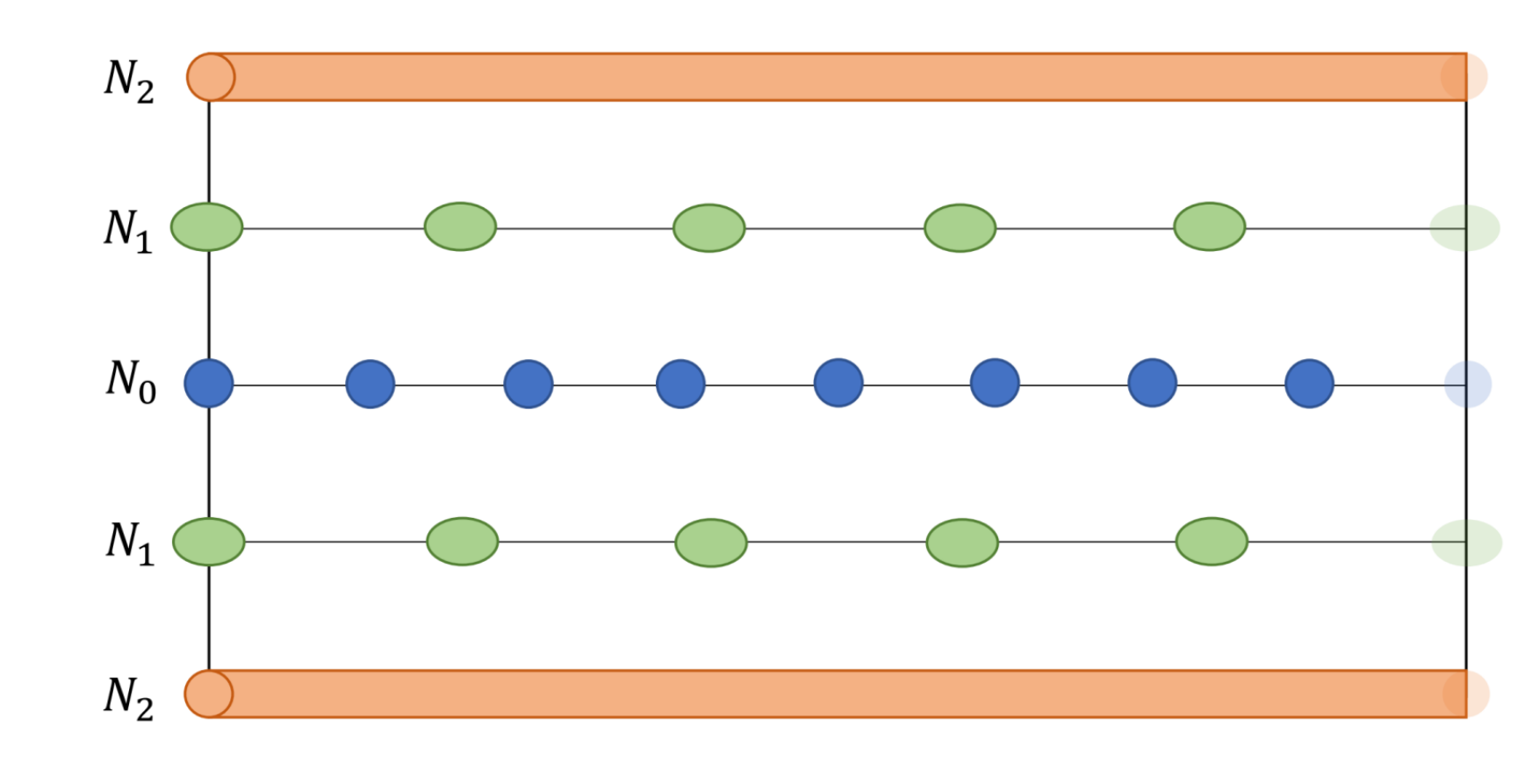}
    }  
    \caption{An example of sampling viewpoints when $N_{0}=8$, (a) sampling on the sphere, (b) sampling on the plane.}
    \label{fig:fig5} 
\end{figure}

Viewpoints of $2\theta, 3\theta, \cdots, \left \lfloor \frac{90}{\theta } \right \rfloor \theta$  degrees north (south) latitude can be sampled by repeating the above procedures. An example is given when $N_{0}=8$ in Fig. \ref{fig:fig5}. In particular, viewpoints of 90 degrees north and south latitude are only sampled once.

\subsubsection{\textbf{Content Weight Calculation}}
\begin{figure}[htbp]
  \centerline{\includegraphics[width=8cm]{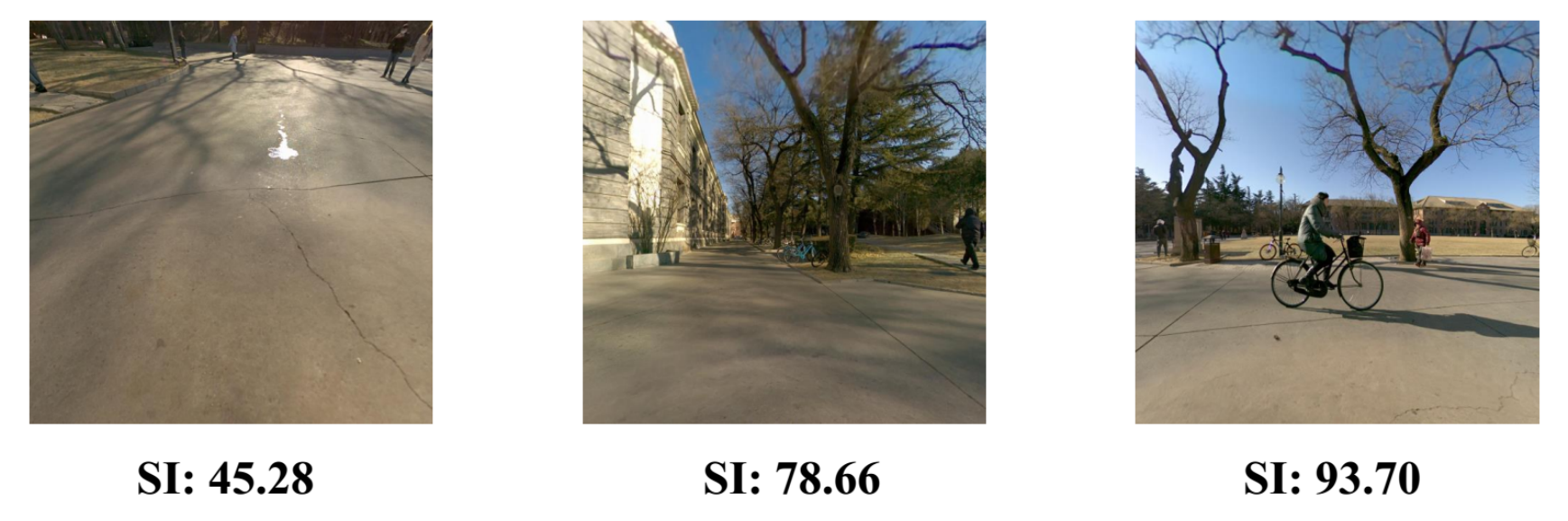}}
  \caption{Viewport images with various SI.}
  \centering
\label{fig:fig6}
\end{figure}

It is a common belief that different regions have different attractions to observers in one image. Regions with salient objects or spatial details tend to catch more attention of viewers which can be reflected by spatial information (SI) \cite{itu1999subjective} as shown in Fig. \ref{fig:fig6}. Higher SI means more details in the viewport image that should be allocated a larger weight. Moreover, the viewport image is three-dimensional,  so we can use the predictive coding based binocular rivalry model to compute the weighted SI of a distorted viewpoint image as follows:
\begin{equation}\label{10}
SI^{Dis}_{L}=std\left [ Sobel\left ( \bm{I}_{L}^{Dis} \right ) \right ],
\end{equation}
\begin{equation}\label{11}
SI^{Dis}_{R}=std\left [ Sobel\left ( \bm{I}_{R}^{Dis} \right ) \right ],
\end{equation}
\begin{equation}\label{12}
CW_{FoVn}=w_{L}SI^{Dis}_{L}+w_{R}SI^{Dis}_{R},
\end{equation}
where $SI_{L}^{Dis}$ and $SI_{R}^{Dis}$ denote the spatial information of distorted left and right viewpoint images $\bm{I}_{L}^{Dis}$ and $\bm{I}_{R}^{Dis}$, $std$ means the standard deviation measurement and $Sobel$ is the Sobel filter. $CW_{FoVn}$ represents the content weight for the $n$-th viewpoint image. $w_{L}$ is the rivalry dominance for left view and so is $w_{R}$. They are given as:
\begin{equation}\label{13}
w_{L}=\frac{1}{N}\sum_{i}v_{i}^{L}EW_{i}^{L}R_{i}^{L},
\end{equation}
\begin{equation}\label{14}
w_{R}=\frac{1}{N}\sum_{i}v_{i}^{R}EW_{i}^{R}R_{i}^{R},
\end{equation}
where $N$ is the number of blocks within the viewport image. $v_{i}$, $EW_{i}$ and $R_{i}$ are described in Section III B.

\subsubsection{\textbf{Location Weight Calculation}}
\begin{figure}[htbp]
  \centerline{\includegraphics[width=7cm]{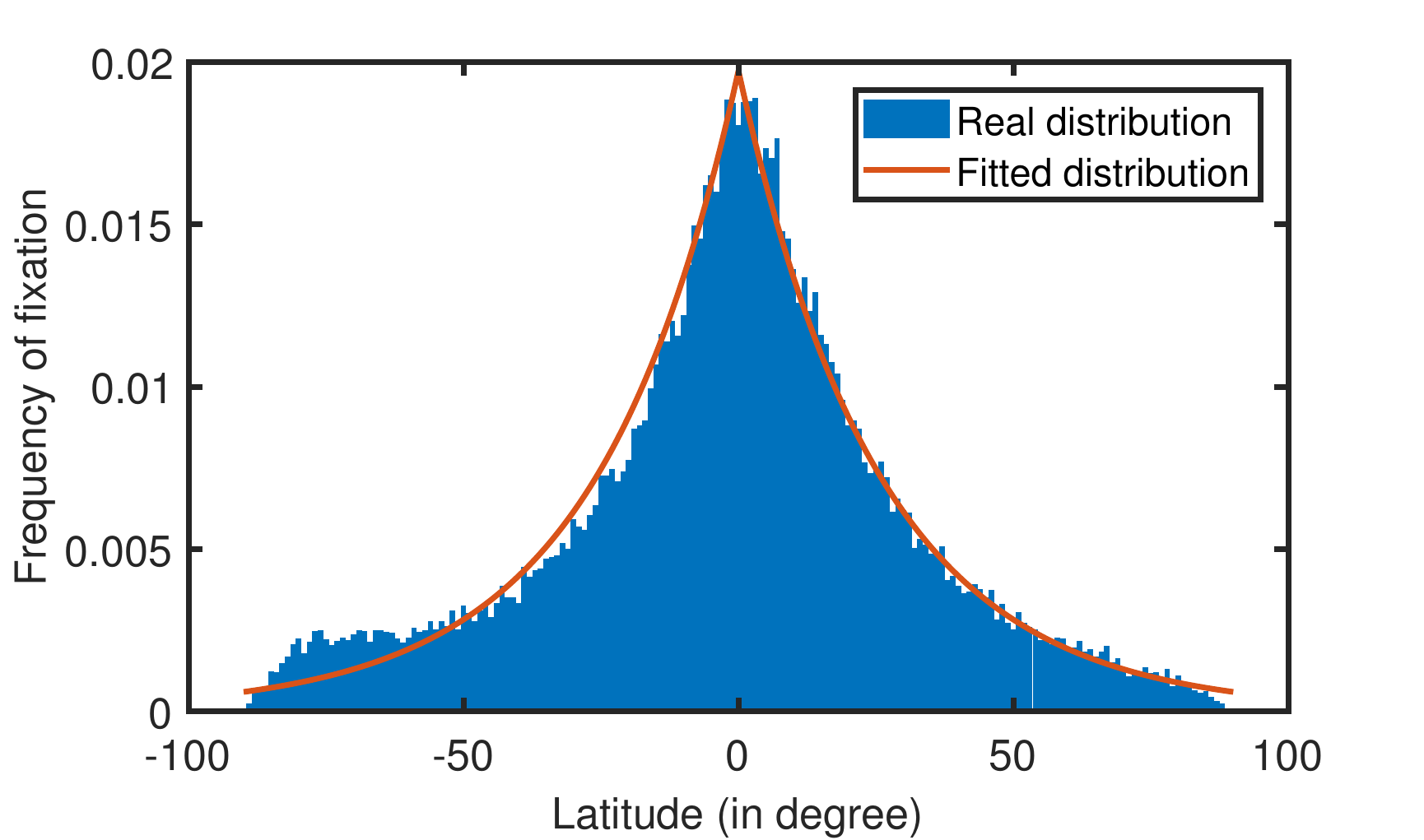}}
  \caption{Viewing frequency versus viewport central latitude.}
  \centering
\label{fig:fig7}
\end{figure}

Subjects tend to view more frequently at the equatorial regions and share a similar possibility of fixating all longitudes in omnidirectional images according to the statistic of the eye-tracking data in a head and eye movements dataset for 360-degree images \cite{rai2017dataset}. In Fig. \ref{fig:fig7}. we model the latitude's probability of being visited using a Laplace distribution based on this dataset. The probability of a viewport image being observed can be regarded as the probability of its viewed central point. Then, the probability of the viewed central point is related to its latitude. Under this assumption, we use the Laplace distribution in Fig. \ref{fig:fig7} to calculate the possibility of one viewpoint image being visited. The location weight (LW) is represented as the viewing probability:
\begin{equation}\label{15}
LW_{FoVn}=P_{FoVn},
\end{equation}
where $LW_{FoVn}$ is the location weight for the $n$-th viewport image and $P_{FoVn}$ represents the probability of the $n$-th observed viewport image. Finally, the weights for $N$ viewport images are normalized to predict the quality score $Q$ of a stereoscopic omnidirectional image as below:

\begin{equation}\label{16}
W_{FoVn}=\frac{CW_{FoVn}LW_{FoVn}}{\sum _{n}CW_{FoVn}LW_{FoVn}},
\end{equation}

\begin{equation}\label{17}
Q=\sum _{n}W_{FoVn}Q_{FoVn},
\end{equation}
 where $W_{FoVn}$ and $Q_{FoVn}$ are the weight and quality of the $n$-th viewpoint image.

\section{Experimental Results and Analysis}
In this section, we conduct experiments on the self-built public stereoscopic omnidirectional image quality assessment database (SOLID) \cite{xu2018subjective} to prove the effectiveness of our proposed metric. The viewport images in our experiment cover a field of view with 90 degrees which is similar to the FoV range when viewing real panoramic images in HMD. Also, $90^{\circ}$ FoV would not bring heavy projection deformation. $N_{0}$ equals $8$ as shown in Fig. \ref{fig:fig5} to keep a balance between computation efficiency and accuracy in viewpoint sampling strategy. In addition, the dictionary size is set as $1024\times256$ to achieve better performance which will be explained in Section IV B. Furthermore, since there are no other available 3D 360-degree image quality databases, the validity of predictive coding based binocular rivalry module and multi-view fusion module are verified on 3D images and 2D panoramic images, respectively.

\begin{figure}[htbp]
    \centering
    \subfigure[]{
        \includegraphics[width=2cm]{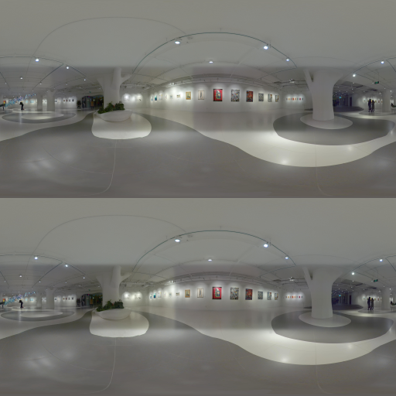}
    }
    \subfigure[]{
        \includegraphics[width=2cm]{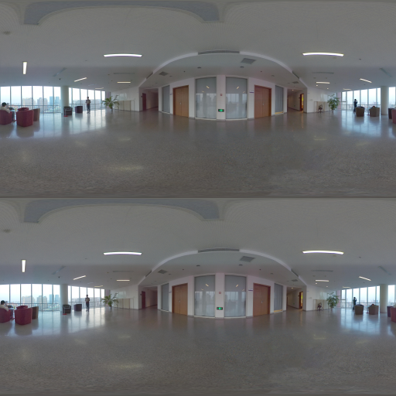}
    }
    \subfigure[]{
        \includegraphics[width=2cm]{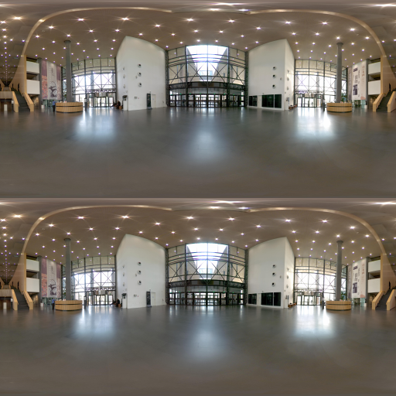}
    }
    \subfigure[]{
        \includegraphics[width=2cm]{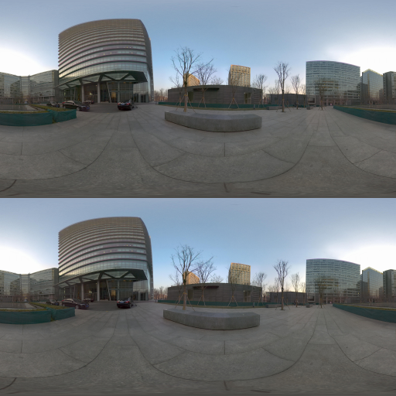}
    }
    \subfigure[]{
        \includegraphics[width=2cm]{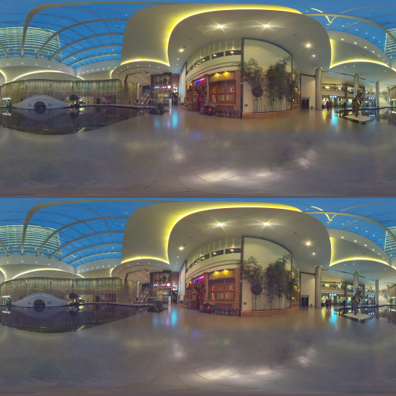}
    }
    \subfigure[]{
        \includegraphics[width=2cm]{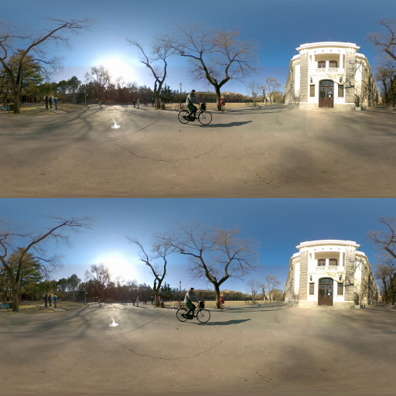}
    }
    \caption{Reference images in SOLID database \cite{xu2018subjective}. The top is the left view image and the bottom is the right view in each reference image. (a) Exihibition (b) Corridor (c) Museum (d) Building (e) Restaurant (f) Road. }
    \label{fig:fig8} 
\end{figure}

\subsection{Databases and Evaluation Measurement}
Image quality assessment databases used in our experiment are listed in Table \ref{table1} and further described in details. 

\begin{table}[htbp]
\centering
\caption{\textsc{Benchmark Test Databases.}}
\label{table1}
\scalebox{0.72}{
\begin{tabular}{@{}cccccc@{}}
\toprule
Database      & \begin{tabular}[c]{@{}c@{}}\# of Source\\ Images\end{tabular} & \begin{tabular}[c]{@{}c@{}}\# of Distorted\\ Images\end{tabular} & \begin{tabular}[c]{@{}c@{}}\# of Distorted\\ Types\end{tabular} & \begin{tabular}[c]{@{}c@{}}Image\\ Type\end{tabular} & \begin{tabular}[c]{@{}c@{}}\# of Assessment\\ Dimentions\end{tabular} \\ \midrule
SOLID \cite{xu2018subjective}         & 6                                                       & 276                                                        & 2                                                         & 3D VR                                                & 3                                                               \\
LIVE Phase I \cite{moorthy2013subjective} & 20                                                      & 365                                                        & 5                                                         & 3D                                                   & 1                                                               \\
LIVE Phase II \cite{chen2013no,chen2013full} & 8                                                       & 360                                                        & 5                                                         & 3D                                                   & 1                                                               \\
OIQA \cite{duan2018perceptual}          & 16                                                      & 320                                                        & 4                                                         & 2D VR                                                & 1                                                               \\
CVIQD2018 \cite{sun2018large}     & 16                                                      & 528                                                        & 3                                                         & 2D VR                                                & 1                                                               \\ \bottomrule
\end{tabular}}
\end{table}

\subsubsection{\textbf{SOLID \cite{xu2018subjective}}}
Our first built stereoscopic omnidirectional image quality assessment database includes 276 distorted images with two distortion types and three depth levels derived from 6 high-quality reference images. Subjective assessments of image quality, depth perception and overall quality are collected in this database. The reference images are impaired by JPEG or BPG compression to simulate image quality degradation. There are 84 symmetrically and 192 asymmetrically distorted images in the database. The corresponding Mean Opinion Scores (MOSs) are provided for the reference image pairs and distorted image pairs. Moreover, the MOS values cover a range of 1 to 5, where higher MOS values represent better image quality. 

\subsubsection{\textbf{LIVE Phase I \cite{moorthy2013subjective}}}
The database is a 3D image database containing 20 original stereoscopic images and 365 symmetrically distorted stereoscopic images. There are five distortion types in this database: Gaussian blurring, additive white noise, JPEG compression, JPEG2000 compression and fast fading for simulating packet loss of transmitted JPEG2000-compressed images. The associated differential Mean Opinion Score (DMOS) which represents human subjective judgments is provided for each stereoscopic image in the range [0, 80]. On the contrary, lower DMOS values mean better visual quality and vice versa.

\subsubsection{\textbf{LIVE Phase II \cite{chen2013no,chen2013full}}}
The stereo image database contains 8 original images and 360 distorted stereoscopic images, which includes 120 symmetrically distorted stereoscopic images and 240 asymmetrically distorted stereoscopic images. The distortion types are the same with that of LIVE Phase I. For each distortion type, every original stereoscopic image is processed to 3 symmetrically distorted stereoscopic images and 6 asymmetrically distorted stereoscopic images. Each distorted stereoscopic image is also associated with a DMOS value evaluated by 33 participants.

\subsubsection{\textbf{OIQA Database \cite{duan2018perceptual}}}
It is a 2D omnidirectional image quality assessment database consisting of 16 pristine images and 320 distorted images under four kinds of distortion types. More specifically, the artifacts include JPEG compression, JPEG2000 compression, Gaussian blur and Gaussian noise. The MOS values are given for both reference and distorted images in the range [1, 10].

\subsubsection{\textbf{CVIQD2018 Database \cite{sun2018large}}}
This 2D panoramic image quality assessment database is the largest compressed 360-degree image database including 16 source images and 528 compressed images. They are compressed with three popular coding technologies, namely JPEG, H.264/AVC and H.265/HEVC. The DMOS values are given for all images in the database.

\subsubsection{\textbf{Performance Measures}}

By following previous methods \cite{brunnstrom2009vqeg}, four commonly used criteria are adopted for quantitative performance evaluation, including Spearman Rank Order Correlation Coefficient (SROCC), Pearson Linear Correlation Coefficient (PLCC), Root Mean Squared Error (RMSE) and Outlier Ratio (OR). SROCC is calculated according to the rank of scores, and it is used to evaluate the prediction monotonicity. PLCC and RMSE are used to evaluate the prediction accuracy. Prediction consistency is given by OR. Higher correlation coefficient means better relevancy with human quality judgements. Lower RMSE and OR means more accurate predictions.
Before evaluating the PLCC, RMSE and OR performance of a quality assessment metric, the logistic mapping is conducted. In this paper, we apply a five-parameter logistic function constrained to monotonic \cite{sheikh2006statistical}:

\begin{equation}\label{18}
y=\beta _{1}\left ( \frac{1}{2}-\frac{1}{1+exp\left ( \beta _{2}\left ( x-\beta _{3} \right ) \right )} \right )+\beta _{4}x+\beta _{5},
\end{equation}

where $x$ denotes the predicted image quality score of the proposed model, $y$ is the corresponding mapped score and $\beta _{i} (i=1,2,3,4,5)$ represent the five parameters which are used to fit the logistic function.

\subsection{Performance Comparison}

Considering our proposed SOIQE is a full-reference (FR) model, to make fair comparison, we compare it with other open source FR state-of-the-art metrics for performance evaluation. It should be noted that the comparison is conducted among several traditional methods rather than learning based methods since the proposed SOIQE is a parametric model without necessary of regressive learning. Moreover, since there is no image quality assessment metric specially designed for stereoscopic omnidirectional images, our proposed model is compared with conventional 2D IQA, 2D OIQA, 3D IQA metrics.

\begin{table}[htbp]
\centering
\caption{\textsc{Performance Evaluation of 12 FR IQA Metrics on SOLID Database \cite{xu2018subjective}. The Best Performing Metric is Highlighted in Bold.}}
\label{table2}
\begin{tabular}{@{}c|c|cccc@{}}
\toprule
Type                     & Metric         & PLCC            & SROCC           & RMSE            & OR              \\ \midrule
\multirow{5}{*}{2D IQA}  & PSNR           & 0.629           & 0.603           & 0.789           & 0.167           \\
                         & SSIM \cite{wang2004image}          & 0.882           &0.888  & 0.478           &0.033  \\
                         & MS-SSIM \cite{wang2003multiscale}        & 0.773           & 0.755           & 0.643           & 0.083           \\
                         & FSIM \cite{zhang2011fsim}          & 0.889           & 0.883           & 0.465           & 0.040           \\
                         & VSI \cite{zhang2014vsi}           & 0.881           & 0.873           & 0.479           &0.033  \\ \midrule
\multirow{3}{*}{2D OIQA} & S-PSNR \cite{yu2015framework}        & 0.593           & 0.567           & 0.816           & 0.188           \\
                         & WS-PSNR \cite{sun2017weighted}       & 0.585           & 0.559           & 0.823           & 0.192           \\
                         & CPP-PSNR \cite{zakharchenko2016quality}      & 0.593           & 0.566           & 0.817           & 0.192           \\ \midrule
\multirow{3}{*}{3D IQA}  & Chen \cite{chen2013full}          & 0.853           & 0.827           & 0.530           & 0.040           \\
                         & W-SSIM \cite{wang2015quality}        & 0.893  & 0.891  & 0.457  & 0.025  \\
                         & W-FSIM \cite{wang2015quality}        & 0.889  & 0.885           & 0.464  & 0.044           \\ \midrule
3D OIQA                  & Proposed SOIQE & \textbf{0.927}  & \textbf{0.924}  & \textbf{0.383}  & \textbf{0.022}  \\ \bottomrule
\end{tabular}
\end{table}

\begin{figure*}[htbp]
    \centering
    \subfigure[]{
        \includegraphics[width=4cm]{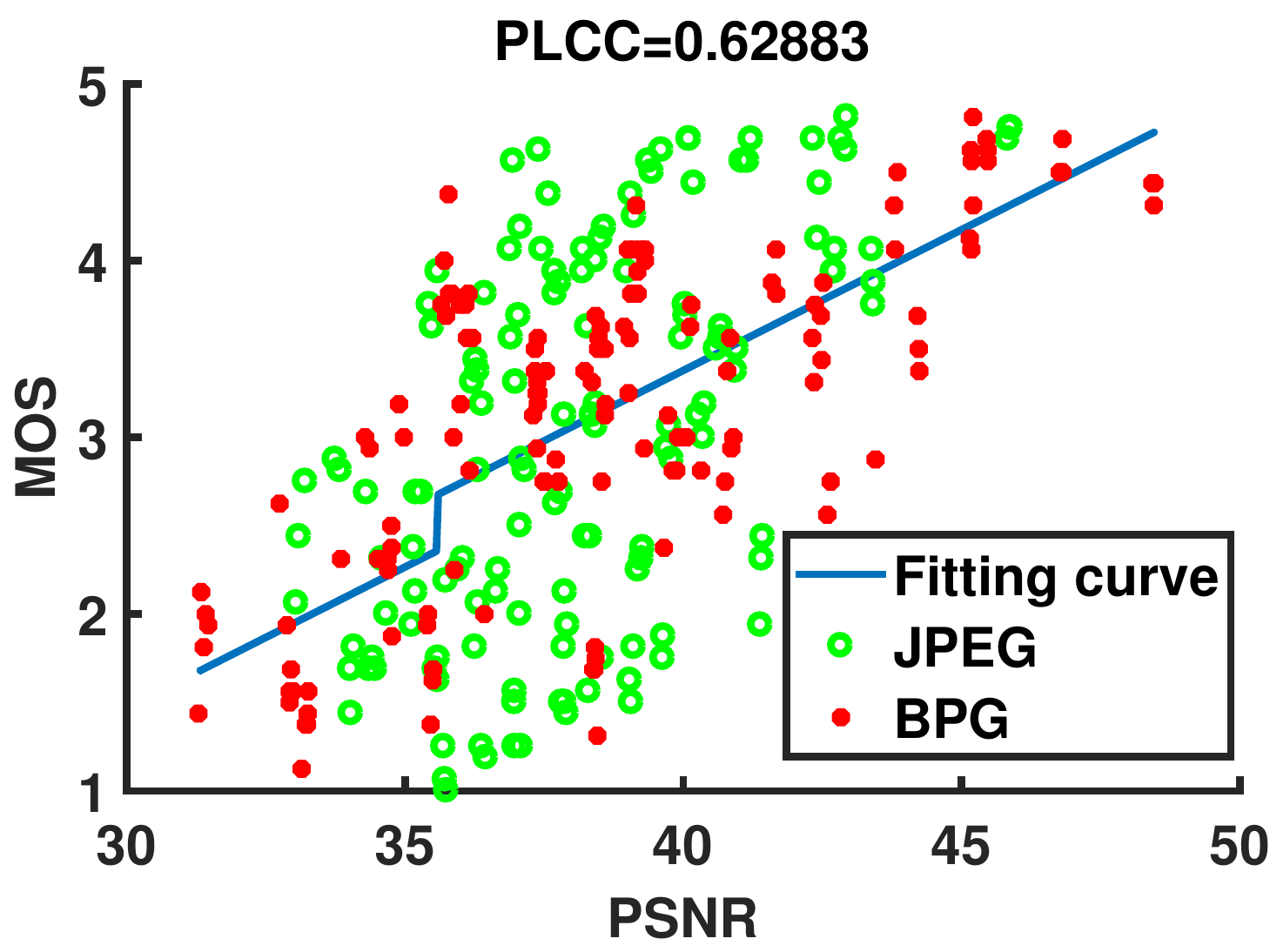}
    }
    \subfigure[]{
        \includegraphics[width=4cm]{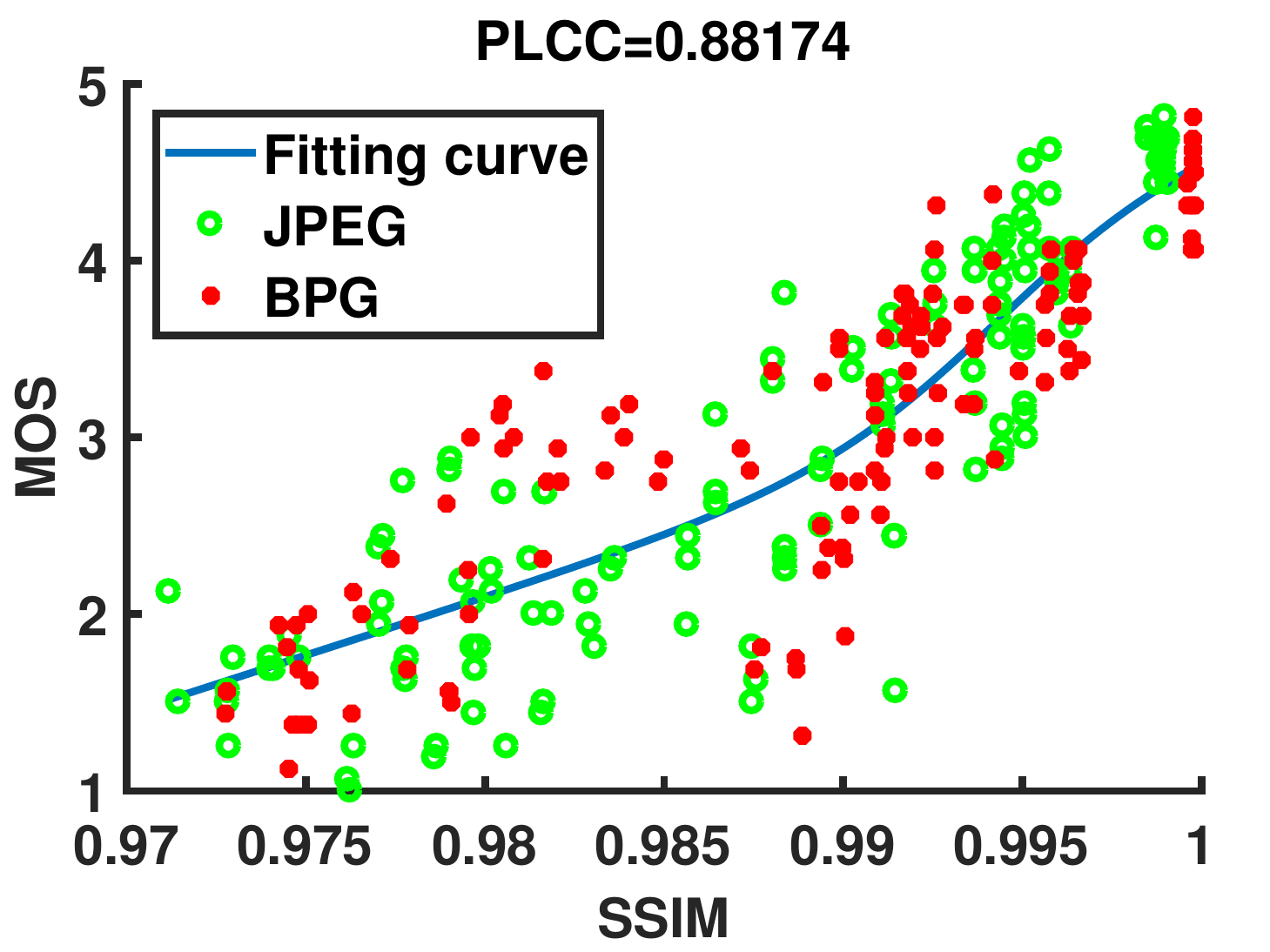}
    }
    \subfigure[]{
        \includegraphics[width=4cm]{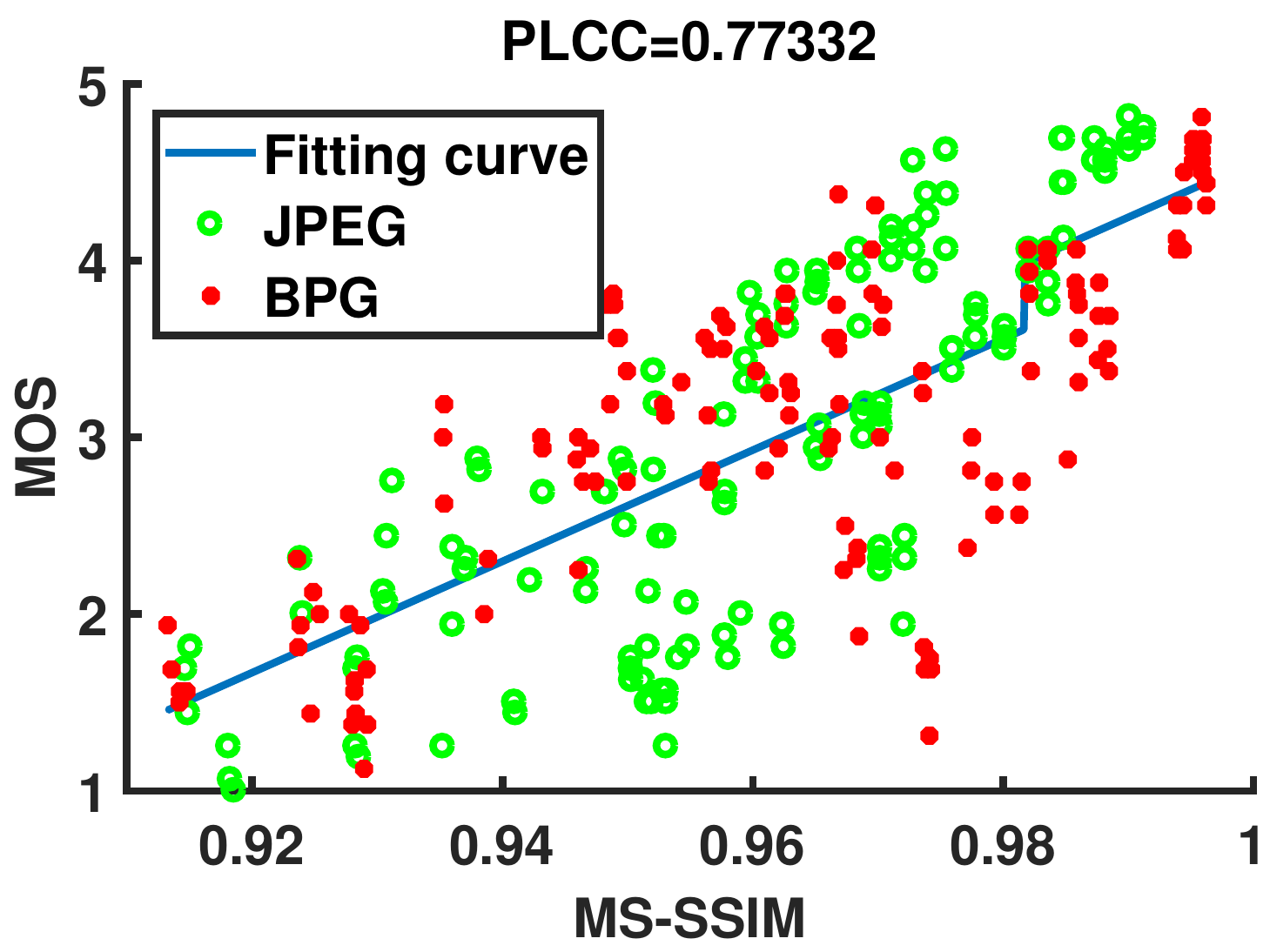}
    }
    \subfigure[]{
        \includegraphics[width=4cm]{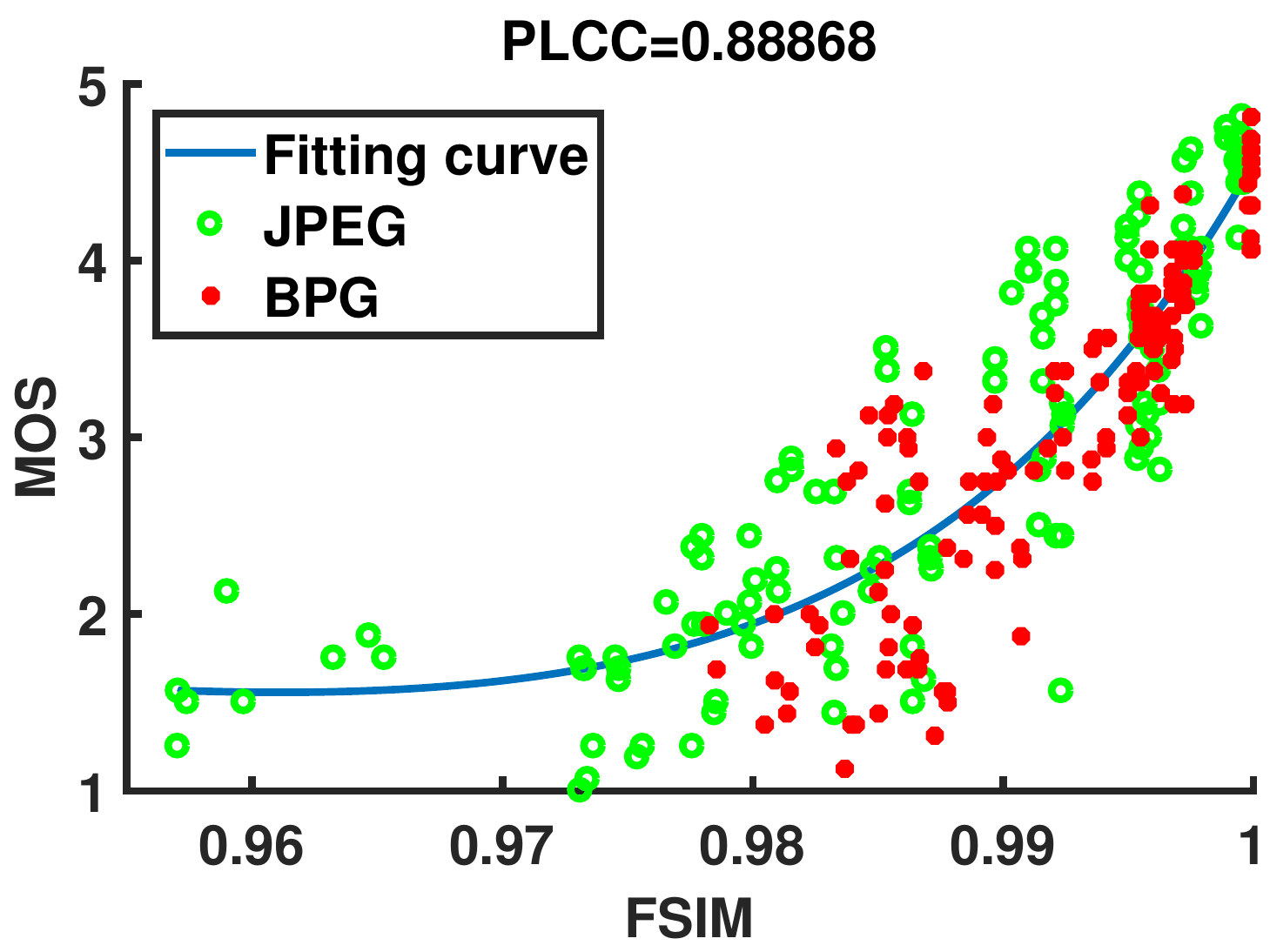}
    }
    \subfigure[]{
        \includegraphics[width=4cm]{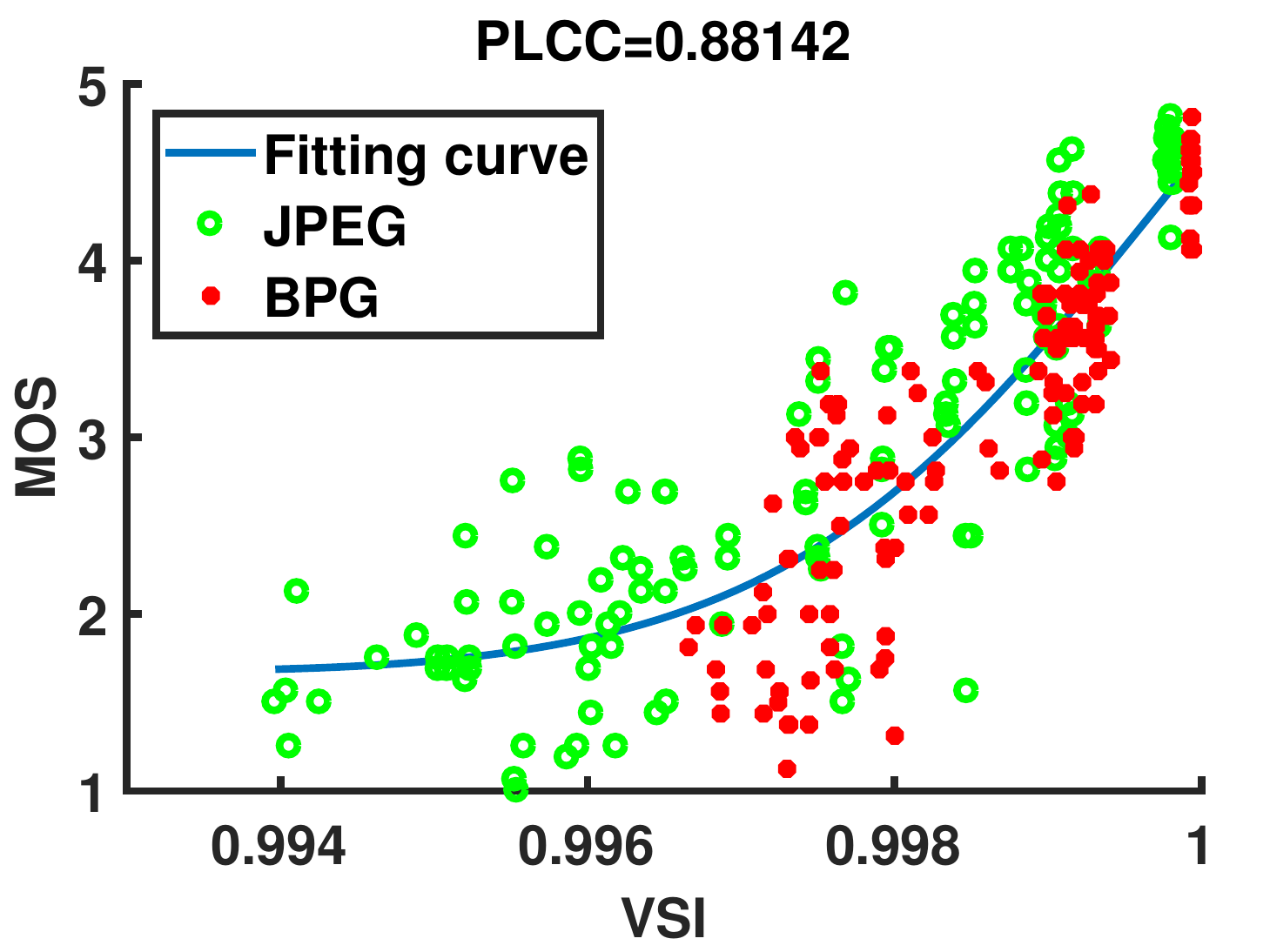}
    }
    \subfigure[]{
        \includegraphics[width=4cm]{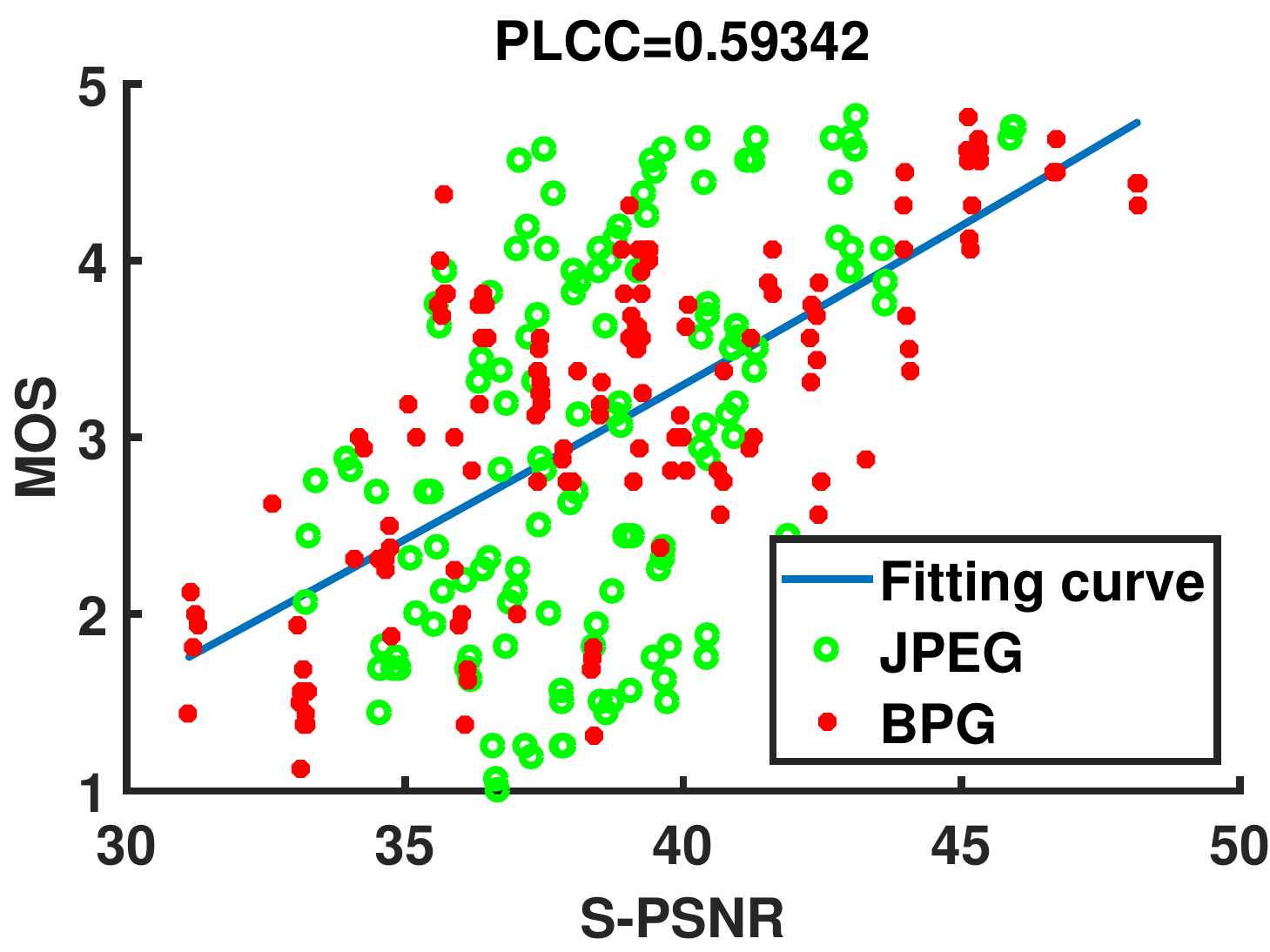}
    }
    \subfigure[]{
        \includegraphics[width=4cm]{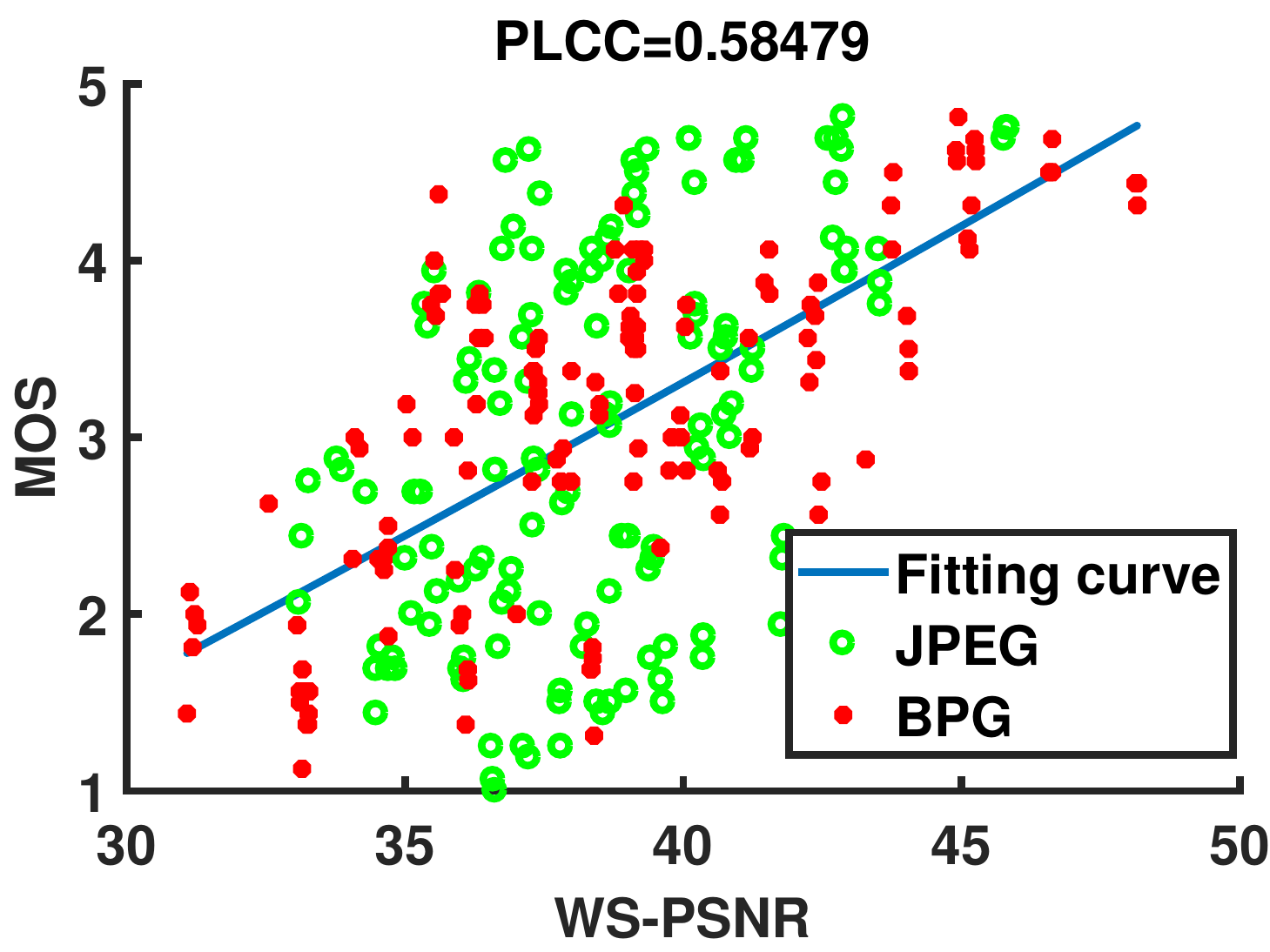}
    }
    \subfigure[]{
        \includegraphics[width=4cm]{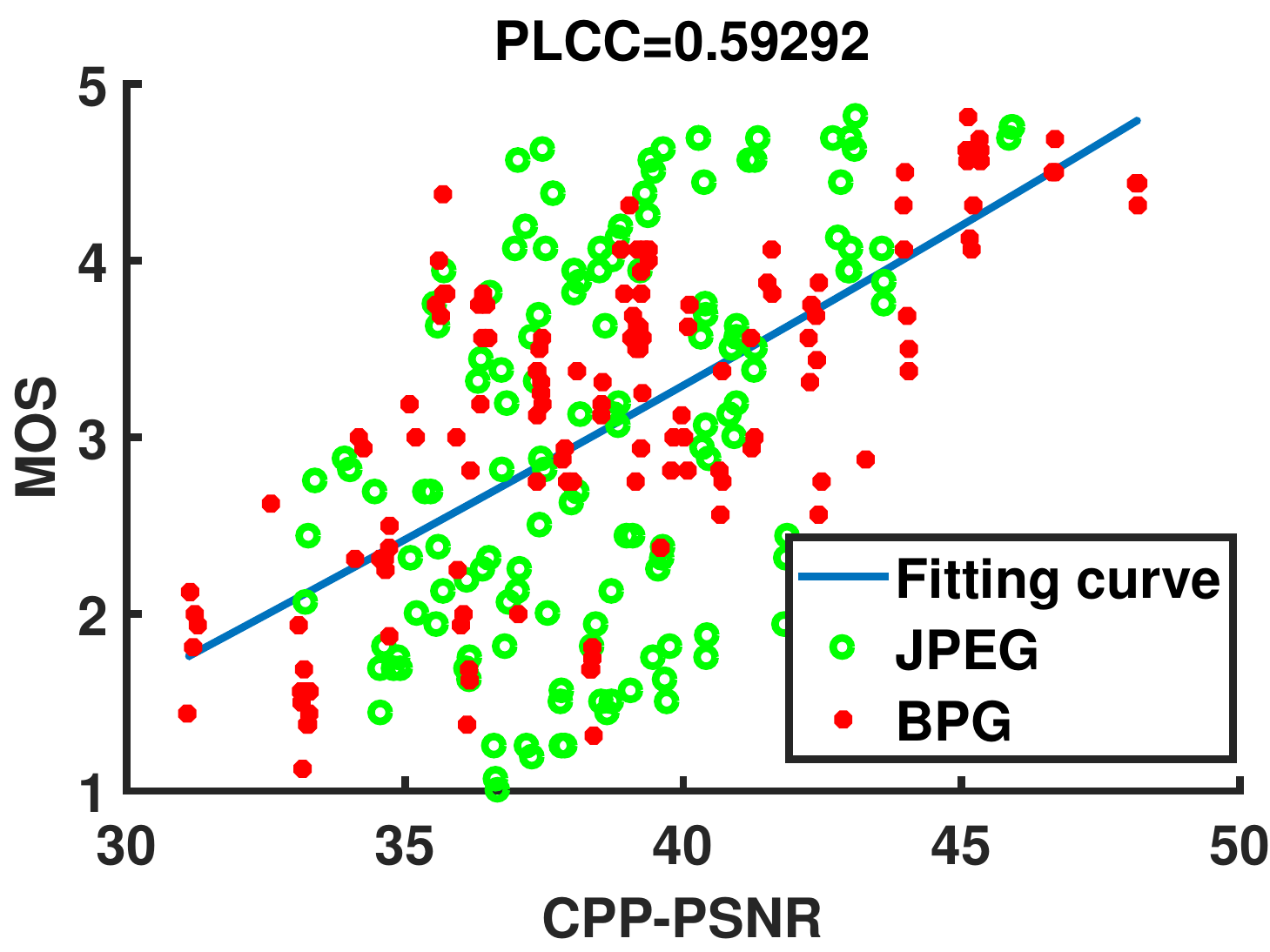}
    }
    \subfigure[]{
        \includegraphics[width=4cm]{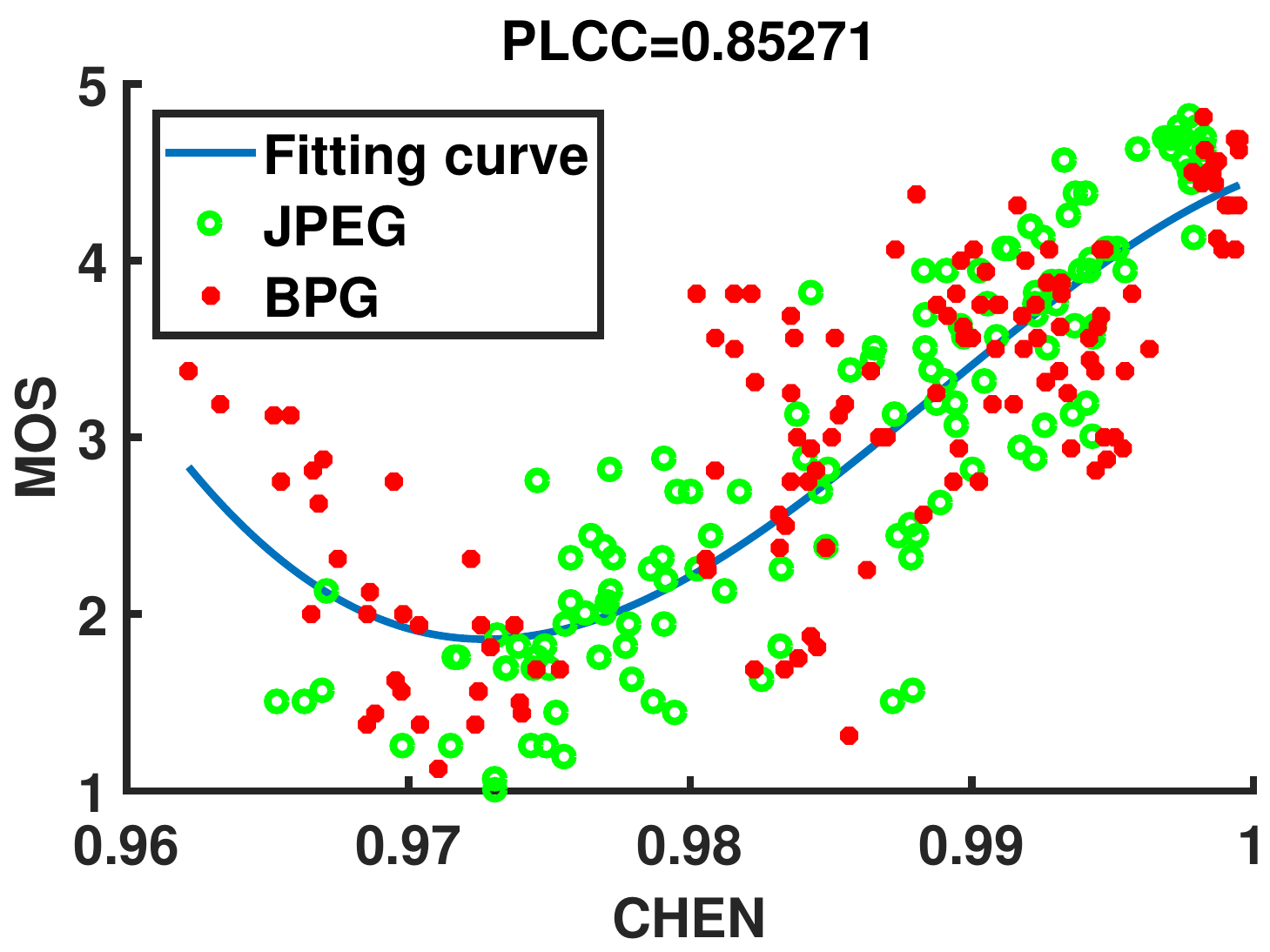}
    }
    \subfigure[]{
        \includegraphics[width=4cm]{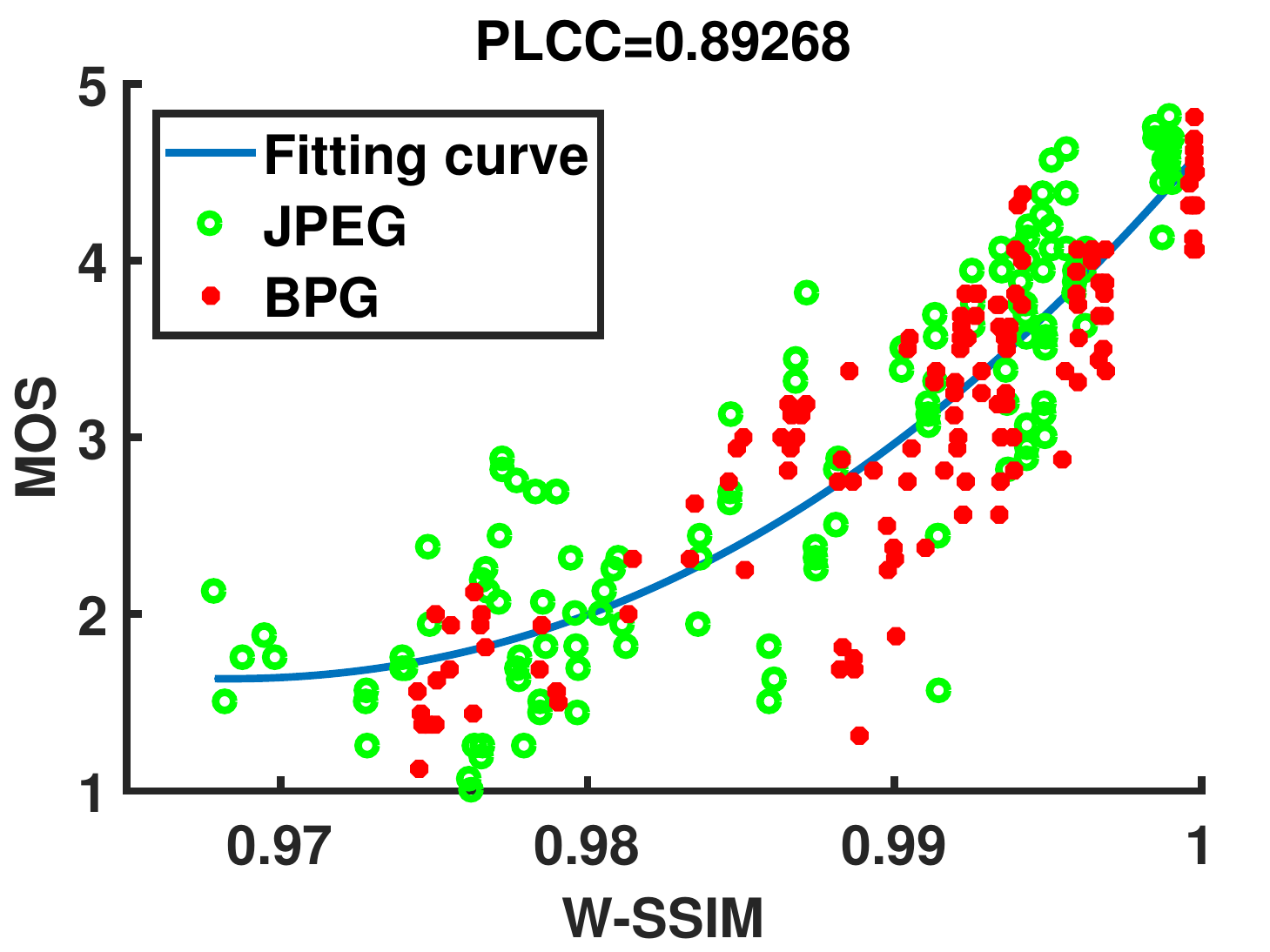}
    }
    \subfigure[]{
        \includegraphics[width=4cm]{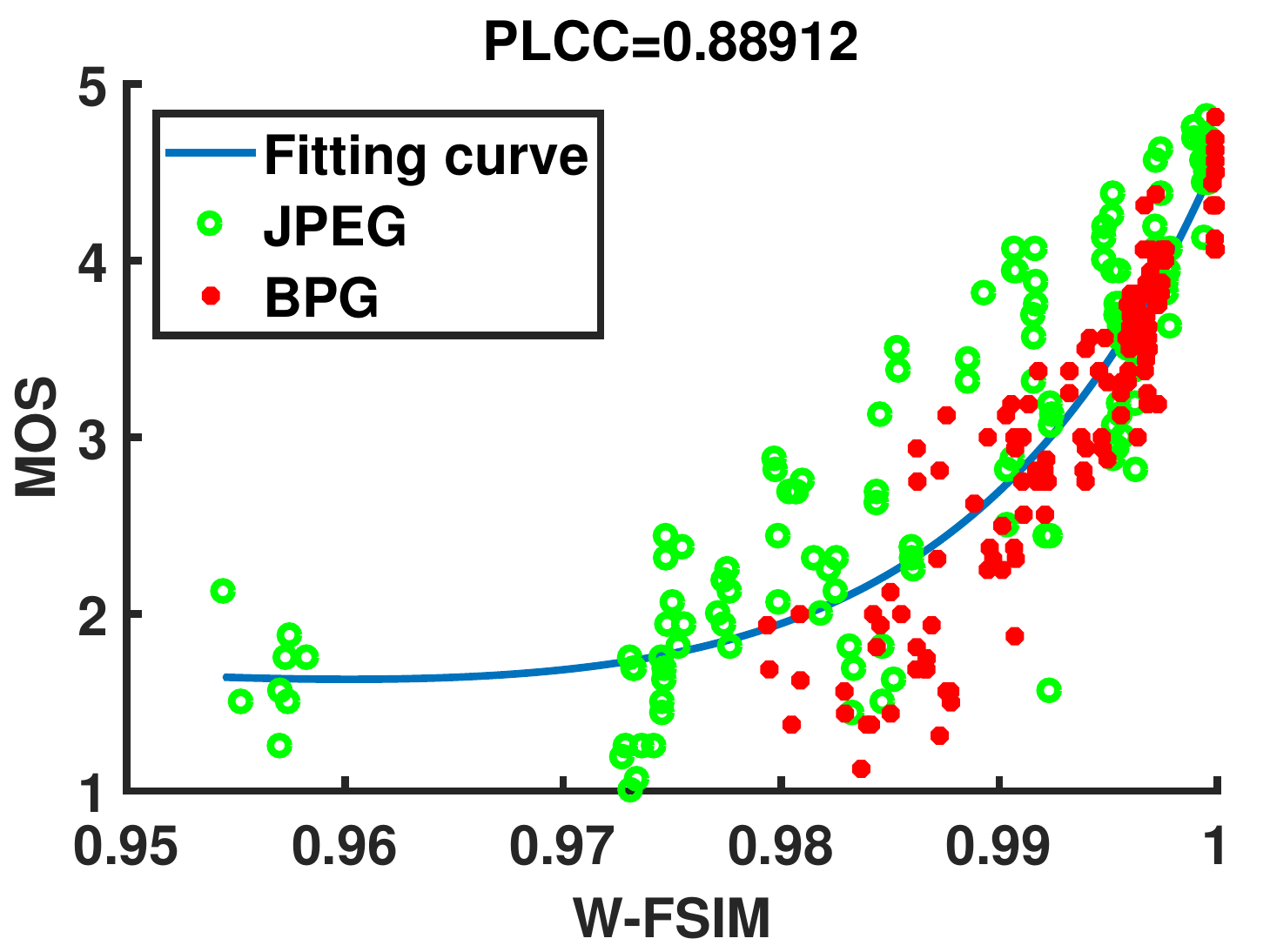}
    }
    \subfigure[]{
        \includegraphics[width=4cm]{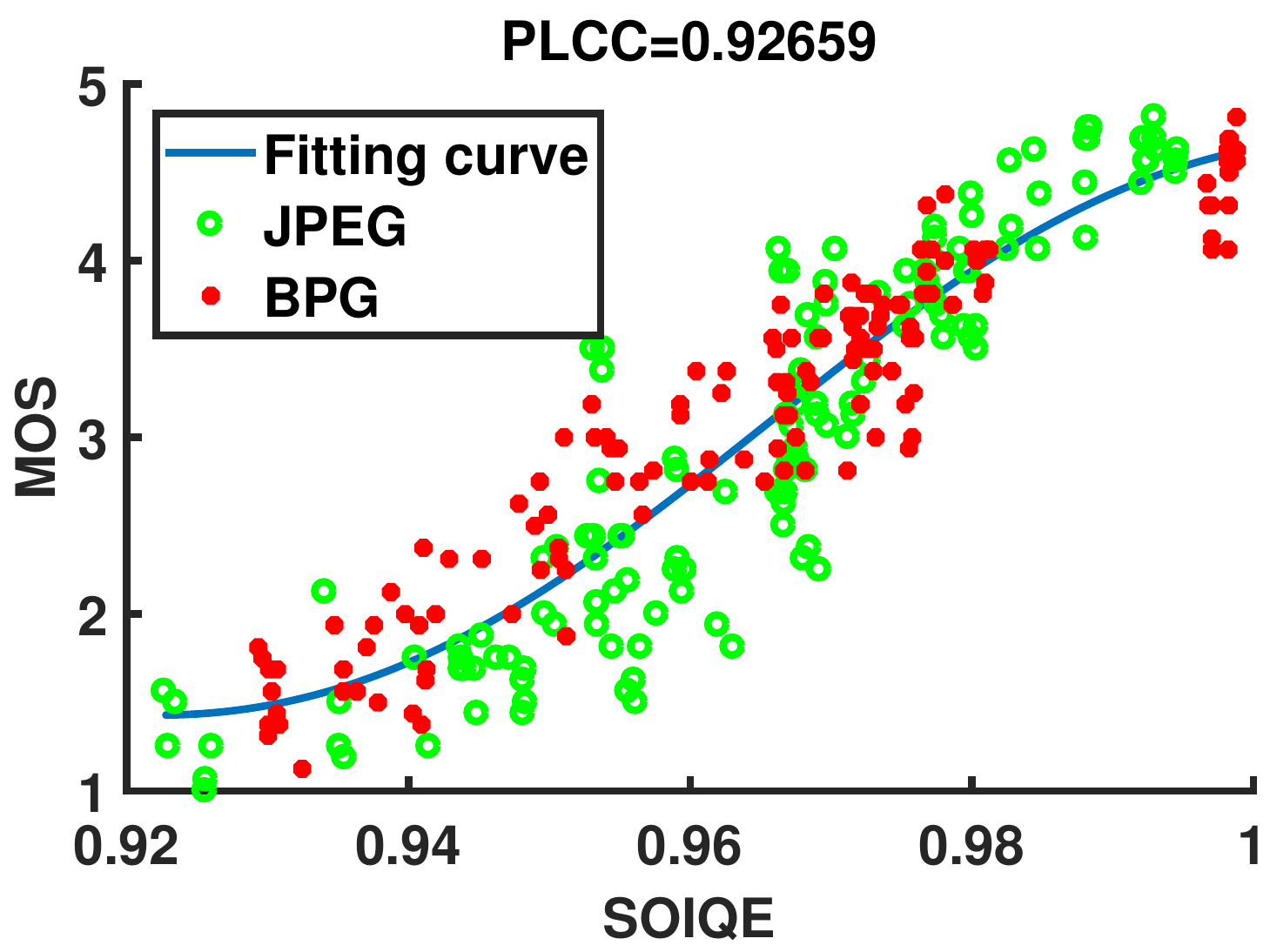}
    }
    \caption{Scatter plots of MOS against predictions by PSNR, SSIM, MS-SSIM, FSIM, VSI, S-PSNR, WS-PSNR, CPP-PSNR, CHEN, W-SSIM, W-FSIM and SOIQE on the SOLID database \cite{xu2018subjective}. Each point indicates one distorted image in the database.}
    \label{fig:fig9} 
\end{figure*} 

Table \ref{table2} shows the PLCC, SROCC, RMSE and OR performance evaluation of 12 FR IQA metrics on SOLID database. These metrics are divided into four types. Firstly, 2D IQA metrics are used for conventional 2D image quality assessment, we choose five commonly used metrics for comparison, namely PSNR, SSIM \cite{wang2004image}, MSSSIM \cite{wang2003multiscale}, FSIM \cite{zhang2011fsim}, and VSI \cite{zhang2014vsi}. The quality scores for left and right view images are averaged to obtain the final predicted score. Besides, 2D OIQA metrics are designed for single view 360-degree image quality assessment. Given that panoramic images are actually viewed on the sphere, three metrics in 360Lib Software \cite{he2016jvet} are utilized for performance measurement including S-PSNR \cite{yu2015framework}, WS-PSNR \cite{sun2017weighted}, and CPP-PSNR \cite{zakharchenko2016quality}. Furthermore, 3D IQA metrics are used to predict the quality of stereoscopic images. Chen \cite{chen2013full} is an open source 3D IQA metric and a weighting policy \cite{wang2015quality} is introduced to further improve the performance of SSIM and FSIM. As listed in Table \ref{table2}, the best performing metric is highlighted in bold and our proposed SOIQE outperforms these state-of-the-art metrics. The proposed SOIQE cares more about the characteristic of stereoscopic omnidirectional images such as eye dominance, FoV range, etc., thus making it a more suitable metric for 3D OIQA.

Apart from the numerical comparison in Table \ref{table2}, scatter plots of MOS values versus the predicted scores of objective metrics are drawn in Fig. \ref{fig:fig9} to give clear and direct results. From this figure, we can see that the predicted scores of SOIQE show better convergency and monotonicity than other metrics ,which means the proposed SOIQE is more accurate.

\subsection{Performance Evaluation on Individual Distortion Type}

To further investigate the differences for individual distortion type, PLCC, SROCC and RMSE performance of the proposed method and other metrics for different distortions are given in Table \ref{table3}. For each kind of distortion, the highest value across the 12 metrics is highlighted in boldface. The proposed metric performs the best both on JPEG and BPG compression distortion. Moreover, the correlations between predicted scores and MOS values of JPEG distortion are usually lower than those of BPG distortion and it can be observed from Table \ref{table3} and Fig. \ref{fig:fig9}. One possible reason is that blocking effects caused by JPEG compression are localized artifacts which may lead to less perceptually separated qualities \cite{chen2013no,moorthy2013subjective}. According to this, IQA of stereoscopic omnidirectional images impaired by JPEG compression seems more challenging in SOLID database.

\begin{table}[htbp]
\centering
\caption{\textsc{Performance Evaluation for Different Distortion Types on SOLID Database \cite{xu2018subjective}. The Best Performing Metric is Highlighted in Bold.}}
\label{table3}
\scalebox{0.95}{
\begin{tabular}{@{}c|cc|cc|cc@{}}
\toprule
               & \multicolumn{2}{c|}{PLCC}                   & \multicolumn{2}{c|}{SROCC}                  & \multicolumn{2}{c}{RMSE}                    \\ \midrule
Metrics        & \multicolumn{1}{c|}{JPEG} & BPG             & \multicolumn{1}{c|}{JPEG} & BPG             & \multicolumn{1}{c|}{JPEG} & BPG             \\ \midrule
PSNR           & 0.564                    & 0.740          & 0.538                    & 0.673          & 0.901                    & 0.624          \\
SSIM \cite{wang2004image}          & 0.907           & 0.857          & 0.893           & 0.879          & 0.460           & 0.477          \\
MS-SSIM \cite{wang2003multiscale}        & 0.841                    & 0.730          & 0.833                    & 0.687          & 0.591                    & 0.633          \\
FSIM \cite{zhang2011fsim}          & 0.894                    & 0.896 & 0.880                    & 0.902 & 0.490                    & 0.411 \\
VSI \cite{zhang2014vsi}           & 0.898                    & 0.888          & 0.885                    & 0.886          & 0.480                    & 0.426          \\ \midrule
SPSNR \cite{yu2015framework}         & 0.515                    & 0.736          & 0.477                    & 0.660          & 0.936                    & 0.627          \\
WSPSNR \cite{sun2017weighted}        & 0.505                    & 0.732          & 0.464                    & 0.658          & 0.949                    & 0.631          \\
CPP-PSNR \cite{zakharchenko2016quality}      & 0.517                    & 0.735          & 0.475                    & 0.660          & 0.934                    & 0.628          \\ \midrule
Chen \cite{chen2013full}          & 0.909           & 0.797          & 0.904           & 0.736          & 0.454           & 0.559          \\
W-SSIM \cite{wang2015quality}        & 0.905                    & 0.887          & 0.888                    & 0.879          & 0.464                    & 0.428          \\
W-FSIM \cite{wang2015quality}        & 0.893                    & 0.933 & 0.885                    & 0.933 & 0.492                    & 0.333 \\ \midrule
Proposed SOIQE & \textbf{0.933}           & \textbf{0.955} & \textbf{0.928}           & \textbf{0.939} & \textbf{0.393}           & \textbf{0.275} \\ \bottomrule
\end{tabular}}
\end{table}

\begin{table}[htbp]
\centering
\caption{\textsc{Performance Evaluation for Symmetrically and Asymmetrically Distorted Images on SOLID Database \cite{xu2018subjective}. The Best Performing Metric is Highlighted in Bold.}}
\label{table4}
\begin{threeparttable}
\scalebox{0.95}{
\begin{tabular}{@{}c|cc|cc|cc@{}}
\toprule
               & \multicolumn{2}{c|}{PLCC}                & \multicolumn{2}{c|}{SROCC}                 & \multicolumn{2}{c}{RMSE}                   \\ \midrule
Metric         & \multicolumn{1}{c|}{Sym} & Asym          & \multicolumn{1}{c|}{Sym} & Asym            & \multicolumn{1}{c|}{Sym} & Asym            \\ \midrule
PSNR           & 0.791                   & 0.394          & 0.789                   & 0.354          & 0.758                   & 0.756          \\
SSIM \cite{wang2004image}           & 0.944          & 0.821          & 0.902          & 0.814          & 0.409          & 0.470          \\
MS-SSIM \cite{wang2003multiscale}        & 0.869                   & 0.631          & 0.836                   & 0.615          & 0.613                   & 0.638          \\
FSIM \cite{zhang2011fsim}          & 0.930                   & 0.853 & 0.890                   & 0.847 & 0.456                   & 0.430 \\
VSI \cite{zhang2014vsi}           & 0.931                   & 0.834          & 0.887                   & 0.807          & 0.454                   & 0.454          \\ \midrule
SPSNR \cite{yu2015framework}         & 0.805                   & 0.364          & 0.766                   & 0.313          & 0.735                   & 0.766          \\
WSPSNR \cite{sun2017weighted}        & 0.807                   & 0.325          & 0.762                   & 0.302          & 0.732                   & 0.778          \\
CPP-PSNR \cite{zakharchenko2016quality}      & 0.806                   & 0.334          & 0.766                   & 0.310          & 0.734                   & 0.775          \\ \midrule
Chen \cite{chen2013full}          & 0.944                   & 0.767          & 0.890                   & 0.700          & 0.411                   & 0.528          \\
W-SSIM \cite{wang2015quality}        & 0.944          & 0.834          & 0.902          & 0.832          & 0.409          & 0.454          \\
W-FSIM \cite{wang2015quality}        & 0.930                   & 0.845 & 0.890                   & 0.842 & 0.456                   & 0.440 \\ \midrule
Proposed SOIQE & \textbf{0.970}          & \textbf{0.867} & \textbf{0.931}          & \textbf{0.866} & \textbf{0.301}          & \textbf{0.411} \\ \bottomrule
\end{tabular}}
 \begin{tablenotes}
        \footnotesize
        \item[1] Sym denotes symmetrical distortion.
        \item[2] Asym denotes asymmetrical distortion.
      \end{tablenotes}
\end{threeparttable}
\vspace{-0.5cm} 
\end{table}

Our SOLID database includes both symmetrical and asymmetrical distortion. Thus, we also validate the performance of SOIQE for symmetrically and asymmetrically distorted stereoscopic panoramic images in Table \ref{table4}. The proposed metric achieves the best performance on both symmetrically and asymmetrically distorted images. Chen \cite{chen2013full} doesn't utilize binocular rivalry model in his algorithm. Although it can perform well on symmetrically distorted stereoscopic images, its performance on asymmetrical distortion is significantly lower than other 3D IQA metrics. Besides, compared with 2D IQA and OIQA metrics, SOIQE shows extraordinary performance on asymmetrical distortion, which further demonstrates the effectiveness of our model.

\subsection{Ablation Study and Parameter Influence}

\begin{figure}[htbp]
  \centerline{\includegraphics[width=6cm]{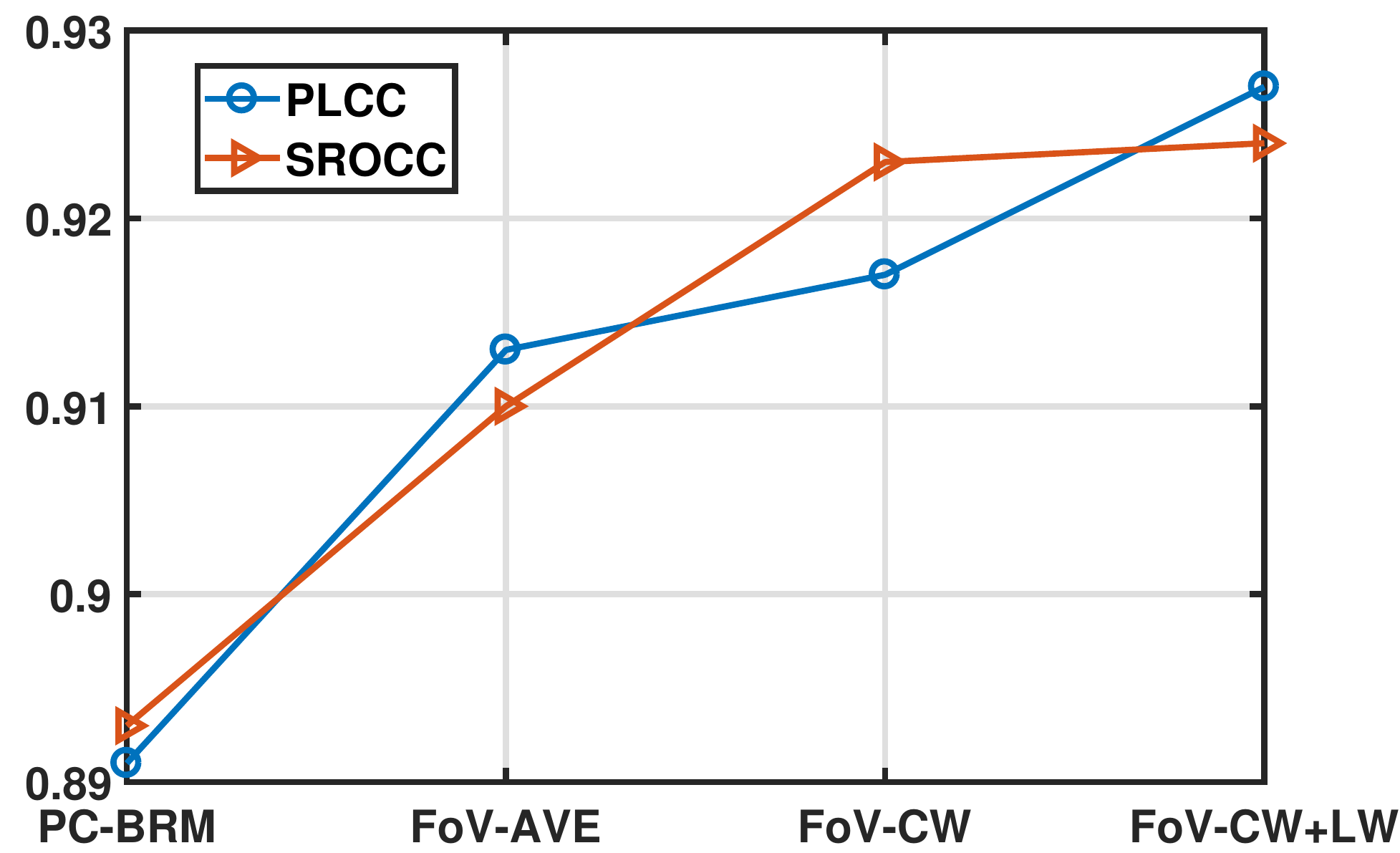}}
  \caption{Performance evaluation of ablation study on SOLID database \cite{xu2018subjective}, where PC-BRM means predictive coding based binocular rivalry model, FoV-AVE represents averaging the quality scores of viewport images, FoV-CW and FoV-CW+LW are aggregating the quality scores using content weight and location weight.}
  \centering
\label{fig:fig10}
\end{figure}

To prove the necessity of every part in our model, ablation study is performed and the results are exhibited in Fig. \ref{fig:fig10}. Firstly, we use the predictive coding based binocular rivalry model to process the entire stereoscopic omnidirectional images. Then, we utilize different weighting policies for aggregating the quality scores of viewport images including averaging, weighting with content alone and further employing location weight. The PLCC performance improves from 0.891 to 0.927 on SOLID database. 

In our model, the dictionary $U=\left [ U_{1},U_{2},\cdots ,U_{k} \right ]$ is constructed by performing gradient descent, where $ U_{i}$ is the $i$-th basis vector (\emph{i.e.}
 pattern) of the dictionary. We examine how the patch size and number of basis vectors affect the performance. In this experiment, we set the patch size as 8, 16 and 32, the number of basis vectors as 512, 1024, 2048 respectively to see the changes of performance. The LIVE IQA database \cite{sheikh2006statistical} is used to generate the dictionary and comparison is performed on SOLID database. The results are presented in Table \ref{table5}. The best performance achieves when patch size equals 16 and the number of basis vectors equals 1024. As shown in Table \ref{table5}, if we achieve better performance, the patch size and number of basis vectors have to increase simultaneously, otherwise, it may suffer from over-fitting or under-fitting problem.

\begin{table}[htbp]
\centering
\caption{\textsc{Performance Evaluation of Different Patch Sizes and Number of Basis Vectors on SOLID database \cite{xu2018subjective}. The Best Three Performing Metrics are Highlighted in Bold.}}
\label{table5}
\begin{threeparttable}
\begin{tabular}{@{}|c|ccc|@{}}
\toprule
\diagbox{PS}{PLCC/SROCC}{SV} & 512         & 1024        & 2048        \\ \midrule
8          & \textbf{0.922/0.924} & 0.882/0.893 & 0.881/0.891 \\
16         & 0.919/0.913 & \textbf{0.927/0.924} & 0.915/0.923 \\
32         & 0.903/0.900 & 0.912/0.910 & \textbf{0.920/0.919} \\ \bottomrule
\end{tabular}
 \begin{tablenotes}
        \footnotesize
        \item[1] PS: Patch sizes.
        \item[2] SV: The number of basis vectors.
      \end{tablenotes}
\end{threeparttable}
\vspace{-0.5cm} 
\end{table}

\subsection{Validation of Predictive Coding Based Binocular Rivalry Model and Multi-view Fusion Model}

Since there is no other available stereoscopic omnidirectional image quality assessment database for validation, we separate the proposed SOIQE into predictive coding based binocular rivalry model and multi-view fusion model. Then, we verify the validity of these two models on public 3D IQA databases and 2D OIQA databases, respectively.

\begin{table}[htbp]
\centering
\caption{\textsc{Performance Evaluation on 3D IQA Database LIVE Phase I \cite{moorthy2013subjective} and II \cite{chen2013no,chen2013full}. The Best Performing Metric is Highlighted in Bold.}}
\label{table6}
\begin{tabular}{@{}c|ccc|ccc@{}}
\toprule
         & \multicolumn{3}{c|}{LIVE Phase I}                & \multicolumn{3}{c}{LIVE Phase II}                \\ \midrule
Metrics  & PLCC           & SROCC          & RMSE           & PLCC           & SROCC          & RMSE           \\ \midrule
You \cite{you2010perceptual}      & 0.830          & 0.814          & 7.746          & 0.800          & 0.786          & 6.772          \\
Benoit \cite{benoit2009quality}  & 0.881          & 0.878          & 7.061 & 0.748          & 0.728          & 7.490          \\
Hewage \cite{hewage2010reduced}  & 0.902 & 0.899 & 9.139          & 0.558          & 0.501          & 9.364          \\
Chen \cite{chen2013full}    & 0.917 & 0.916 & 6.533 & 0.900 & 0.889 & 4.987 \\
Chen \cite{chen2013no}    & 0.895          & 0.891          & 7.247          & 0.895 &0.880 & 5.102 \\
Bensalma  \cite{bensalma2013perceptual}& 0.887          & 0.875          & 7.559          & 0.770          & 0.751          & 7.204          \\
Proposed & \textbf{0.920} & \textbf{0.917} & \textbf{6.266} & \textbf{0.915} & \textbf{0.907} & \textbf{4.544} \\ \bottomrule
\end{tabular}
\end{table}

The performance of predictive coding based binocular rivalry model on LIVE 3D IQA databases Phase I and II are presented in Table \ref{table6}. We compare this model with some well-known stereoscopic IQA algorithms. The best result is highlighted in boldface. From Table \ref{table6}, we can see that the proposed metric achieves the best performance on both LIVE Phase-I and Phase-II databases. There is asymmetrical distortion on LIVE Phase-II database, thus it demonstrates the effectiveness of proposed predictive coding based binocular rivalry model that outperforms other metrics which are based on conventional binocular rivalry model on LIVE Phase-II database. In general, the proposed metric correlates much more consistently with subjective evaluations than other metrics.

\begin{table}[]
\centering
\caption{\textsc{Performance Evaluation on 2D VR IQA Database OIQA \cite{duan2018perceptual} and CVIQD2018 \cite{sun2018large}.}}
\label{table7}
\begin{threeparttable}
\scalebox{0.9}{
\begin{tabular}{@{}ccccccc@{}}
\toprule
\multicolumn{7}{c}{OIQA Database} \\ \midrule
\multicolumn{1}{c|}{} & \multicolumn{3}{c|}{PLCC} & \multicolumn{3}{c}{SROCC} \\ \midrule
\multicolumn{1}{c|}{Metrics} & \multicolumn{1}{c|}{Orignal} & \multicolumn{1}{c|}{MvFM} & \multicolumn{1}{c|}{Gain} & \multicolumn{1}{c|}{Orignal} & \multicolumn{1}{c|}{MvFM} & Gain \\ \midrule
\multicolumn{1}{c|}{PSNR} & 0.509 & 0.592 & \multicolumn{1}{c|}{+16.31\%} & 0.498 & 0.579 & +16.27\% \\
\multicolumn{1}{c|}{SSIM \cite{wang2004image}} & 0.882 & 0.884 & \multicolumn{1}{c|}{+0.23\%} & 0.871 & 0.873 & +0.23\% \\
\multicolumn{1}{c|}{MS-SSIM \cite{wang2003multiscale}} & 0.677 & 0.713 & \multicolumn{1}{c|}{+5.32\%} & 0.666 & 0.705 & +5.86\% \\
\multicolumn{1}{c|}{FSIM \cite{zhang2011fsim}} & 0.917 & 0.931 & \multicolumn{1}{c|}{+1.53\%} & 0.911 & 0.926 & +1.65\% \\
\multicolumn{1}{c|}{VSI \cite{zhang2014vsi}} & 0.906 & 0.926 & \multicolumn{1}{c|}{+2.21\%} & 0.902 & 0.920 & +2.00\% \\ \midrule
\multicolumn{1}{c|}{Average} & - & - & \multicolumn{1}{c|}{+5.12\%} & - & - & +5.20\% \\ \midrule
\multicolumn{7}{c}{CVIQD2018 Database} \\ \midrule
\multicolumn{1}{c|}{} & \multicolumn{3}{c|}{PLCC} & \multicolumn{3}{c}{SROCC} \\ \midrule
\multicolumn{1}{c|}{Metrics} & \multicolumn{1}{c|}{Orignal} & \multicolumn{1}{c|}{MvFM} & \multicolumn{1}{c|}{Gain} & \multicolumn{1}{c|}{Orignal} & \multicolumn{1}{c|}{MvFM} & Gain \\ \midrule
\multicolumn{1}{c|}{PSNR} & 0.751 & 0.840 & \multicolumn{1}{c|}{+11.85\%} & 0.729 & 0.827 & +13.44\% \\
\multicolumn{1}{c|}{SSIM \cite{wang2004image}} & 0.818 & 0.889 & \multicolumn{1}{c|}{+8.68\%} & 0.832 & 0.889 & +6.85\% \\
\multicolumn{1}{c|}{MS-SSIM \cite{wang2003multiscale}} & 0.809 & 0.893 & \multicolumn{1}{c|}{+10.38\%} & 0.820 & 0.891 & +8.66\% \\
\multicolumn{1}{c|}{FSIM \cite{zhang2011fsim}} & 0.882 & 0.911 & \multicolumn{1}{c|}{+3.29\%} & 0.880 & 0.902 & +2.50\% \\
\multicolumn{1}{c|}{VSI \cite{zhang2014vsi}} & 0.907 & 0.919 & \multicolumn{1}{c|}{+1.32\%} & 0.900 & 0.908 & +0.89\% \\ \cmidrule(l){2-7} 
\multicolumn{1}{c|}{Average} & - & - & \multicolumn{1}{c|}{+7.10\%} & - & - & +6.47\% \\ \bottomrule
\end{tabular}}
 \begin{tablenotes}
        \footnotesize
        \item[1] MvFM: Multi-view Fusion Model.
        \item[2] Gain: The performance improved compared with Original.
      \end{tablenotes}
\end{threeparttable}
\vspace{-0.5cm} 
\end{table}

Moreover, we conduct experiments on two panoramic image quality assessment databases and the performance results are demonstrated in Table \ref{table7}. Note that CVIQD2018 database includes 16 pristine omnidirectional images, twelve of which are shot on the ground while the other four are captured in the air. The probability distribution of viewing directions may differ in these two kinds of photos. However, the location weight in our model is calculated from the empirical distribution acquired from the head and eye movements dataset \cite{rai2017dataset} in which the images are almost shot on the ground. Consequently, we exclude four reference pictures taken in the air and there remain 12 reference images and 396 distorted images in CVIQD2018 database. In Table \ref{table7}, five FR metrics PSNR, SSIM \cite{wang2004image}, MS-SSIM \cite{wang2003multiscale}, FSIM \cite{zhang2011fsim} and VSI \cite{zhang2014vsi} are tested on the panoramic images, then these metrics are optimized with multi-view fusion model (MvFM) and their performance on 2D omnidirectional images has been improved, which proves the effectiveness of MvFM.

\section{Conclusions}

In this paper, to solve the challenging problem SOIQA, we propose SOIQE that contains the predictive coding based binocular rivalry module and the multi-view fusion module. The predictive coding based binocular rivalry model is inspired by the  HVS. It holds the point that it is the competition between high-level patterns that plays a significant role in rivalry dominance. To the best of our knowledge, it is the very first work to introduce predictive coding theory into modeling binocular rivalry in SIQA as well as SOIQA. Moreover, we present a multi-view fusion model for aggregating the quality scores of viewport images. Content weight and location weight are derived from users’ preference for scene contents and viewing directions. Several state-of-the-art 2D/3D IQA and 2D OIQA metrics are compared with our model on five public databases. Experiment results show that SOIQE has excellent ability for predicting the visual quality of stereoscopic omnidirectional images for both symmetrically and asymmetrically degraded images of various distortion types. Besides, it outperforms the classic metrics both on 3D images and 2D panoramic images which verifies its generalization and robustness.

In the future, we believe a deep insight on the likelihood and prior in binocular rivalry will be beneficial to 3D omnidirectional video quality assessment. In addition, reference images are usually unavailable in real situations, we will develop a no-reference SOIQA metric for better practical applications in future research. Moreover, apart from image quality, it is significant to understand human perception on other dimensions such as depth perception, visual comfort, overall quality of experience to further improve the user experience of stereoscopic omnidirectional images.

\bibliographystyle{IEEEtran}
\bibliography{references}

\begin{thebibliography}{10}
\providecommand{\url}[1]{#1}
\csname url@samestyle\endcsname
\providecommand{\newblock}{\relax}
\providecommand{\bibinfo}[2]{#2}
\providecommand{\BIBentrySTDinterwordspacing}{\spaceskip=0pt\relax}
\providecommand{\BIBentryALTinterwordstretchfactor}{4}
\providecommand{\BIBentryALTinterwordspacing}{\spaceskip=\fontdimen2\font plus
\BIBentryALTinterwordstretchfactor\fontdimen3\font minus
  \fontdimen4\font\relax}
\providecommand{\BIBforeignlanguage}[2]{{%
\expandafter\ifx\csname l@#1\endcsname\relax
\typeout{** WARNING: IEEEtran.bst: No hyphenation pattern has been}%
\typeout{** loaded for the language `#1'. Using the pattern for}%
\typeout{** the default language instead.}%
\else
\language=\csname l@#1\endcsname
\fi
#2}}
\providecommand{\BIBdecl}{\relax}
\BIBdecl

\bibitem{vrwhitepaper2017}
X.~Chen, ``Virtual reality/augmented reality white paper,'' 2017.

\bibitem{diemer2015impact}
J.~Diemer, G.~W. Alpers, H.~M. Peperkorn, Y.~Shiban, and A.~M{\"u}hlberger,
  ``The impact of perception and presence on emotional reactions: a review of
  research in virtual reality,'' \emph{Frontiers in psychology}, vol.~6, p.~26,
  2015.

\bibitem{cabral2016introducing}
B.~K. Cabral, ``Introducing {Facebook Surround 360}: An open, high-quality
  {3D-360} video capture system. 2016,'' 2016.

\bibitem{wang2006modern}
Z.~Wang and A.~C. Bovik, ``Modern image quality assessment,'' \emph{Synthesis
  Lectures on Image, Video, and Multimedia Processing}, vol.~2, no.~1, pp.
  1--156, 2006.

\bibitem{seshadrinathan2010study}
K.~Seshadrinathan, R.~Soundararajan, A.~C. Bovik, and L.~K. Cormack, ``Study of
  subjective and objective quality assessment of video,'' \emph{IEEE
  transactions on Image Processing}, vol.~19, no.~6, pp. 1427--1441, 2010.

\bibitem{ERP}
``Equirectangular projection,''
  \url{https://en.wikipedia.org/wiki/Equirectangular_projection}.

\bibitem{greene1986environment}
N.~Greene, ``Environment mapping and other applications of world projections,''
  \emph{IEEE Computer Graphics and Applications}, vol.~6, no.~11, pp. 21--29,
  1986.

\bibitem{chen2018blind}
Z.~Chen, W.~Zhou, and W.~Li, ``Blind stereoscopic video quality assessment:
  From depth perception to overall experience,'' \emph{IEEE Transactions on
  Image Processing}, vol.~27, no.~2, pp. 721--734, 2018.

\bibitem{tam2011stereoscopic}
W.~J. Tam, F.~Speranza, S.~Yano, K.~Shimono, and H.~Ono, ``Stereoscopic
  {3D-TV}: visual comfort,'' \emph{IEEE Transactions on Broadcasting}, vol.~57,
  no.~2, pp. 335--346, 2011.

\bibitem{kim2019vrsa}
H.~G. Kim, H.-T. Lim, S.~Lee, and Y.~M. Ro, ``{VRSA Net:} vr sickness
  assessment considering exceptional motion for 360бу vr video,'' \emph{IEEE
  Transactions on Image Processing}, vol.~28, no.~4, pp. 1646--1660, 2019.

\bibitem{wang2004image}
Z.~Wang, A.~C. Bovik, H.~R. Sheikh, E.~P. Simoncelli \emph{et~al.}, ``Image
  quality assessment: from error visibility to structural similarity,''
  \emph{IEEE transactions on image processing}, vol.~13, no.~4, pp. 600--612,
  2004.

\bibitem{wang2003multiscale}
Z.~Wang, E.~P. Simoncelli, and A.~C. Bovik, ``Multiscale structural similarity
  for image quality assessment,'' in \emph{The Thrity-Seventh Asilomar
  Conference on Signals, Systems \& Computers, 2003}, vol.~2.\hskip 1em plus
  0.5em minus 0.4em\relax Ieee, 2003, pp. 1398--1402.

\bibitem{zhang2011fsim}
L.~Zhang, L.~Zhang, X.~Mou, and D.~Zhang, ``{FSIM}: A feature similarity index
  for image quality assessment,'' \emph{IEEE transactions on Image Processing},
  vol.~20, no.~8, pp. 2378--2386, 2011.

\bibitem{yasakethu2008quality}
S.~Yasakethu, C.~T. Hewage, W.~A.~C. Fernando, and A.~M. Kondoz, ``Quality
  analysis for {3D} video using {2D} video quality models,'' \emph{IEEE
  Transactions on Consumer Electronics}, vol.~54, no.~4, pp. 1969--1976, 2008.

\bibitem{pinson2004new}
M.~H. Pinson and S.~Wolf, ``A new standardized method for objectively measuring
  video quality,'' \emph{IEEE Transactions on broadcasting}, vol.~50, no.~3,
  pp. 312--322, 2004.

\bibitem{benoit2009quality}
A.~Benoit, P.~Le~Callet, P.~Campisi, and R.~Cousseau, ``Quality assessment of
  stereoscopic images,'' \emph{EURASIP journal on image and video processing},
  vol. 2008, no.~1, p. 659024, 2009.

\bibitem{you2010perceptual}
J.~You, L.~Xing, A.~Perkis, and X.~Wang, ``Perceptual quality assessment for
  stereoscopic images based on {2D} image quality metrics and disparity
  analysis,'' in \emph{Proc. Int. Workshop Video Process. Quality Metrics
  Consum. Electron}, vol.~9, 2010, pp. 1--6.

\bibitem{levelt1965binocular}
W.~J. Levelt, ``On binocular rivalry,'' Ph.D. dissertation, Van Gorcum Assen,
  1965.

\bibitem{ohzawa1998mechanisms}
I.~Ohzawa, ``Mechanisms of stereoscopic vision: the disparity energy model,''
  \emph{Current opinion in neurobiology}, vol.~8, no.~4, pp. 509--515, 1998.

\bibitem{ding2006gain}
J.~Ding and G.~Sperling, ``A gain-control theory of binocular combination,''
  \emph{Proceedings of the National Academy of Sciences}, vol. 103, no.~4, pp.
  1141--1146, 2006.

\bibitem{ryu2014no}
S.~Ryu and K.~Sohn, ``No-reference quality assessment for stereoscopic images
  based on binocular quality perception,'' \emph{IEEE Transactions on Circuits
  and Systems for Video Technology}, vol.~24, no.~4, pp. 591--602, 2014.

\bibitem{chen2013full}
M.-J. Chen, C.-C. Su, D.-K. Kwon, L.~K. Cormack, and A.~C. Bovik,
  ``Full-reference quality assessment of stereopairs accounting for rivalry,''
  \emph{Signal Processing: Image Communication}, vol.~28, no.~9, pp.
  1143--1155, 2013.

\bibitem{chen2013no}
M.-J. Chen, L.~K. Cormack, and A.~C. Bovik, ``No-reference quality assessment
  of natural stereopairs,'' \emph{IEEE Transactions on Image Processing},
  vol.~22, no.~9, pp. 3379--3391, 2013.

\bibitem{smola2004tutorial}
A.~J. Smola and B.~Sch{\"o}lkopf, ``A tutorial on support vector regression,''
  \emph{Statistics and computing}, vol.~14, no.~3, pp. 199--222, 2004.

\bibitem{lin2014quality}
Y.-H. Lin and J.-L. Wu, ``Quality assessment of stereoscopic {3D} image
  compression by binocular integration behaviors,'' \emph{IEEE transactions on
  Image Processing}, vol.~23, no.~4, pp. 1527--1542, 2014.

\bibitem{wang2015quality}
J.~Wang, A.~Rehman, K.~Zeng, S.~Wang, and Z.~Wang, ``Quality prediction of
  asymmetrically distorted stereoscopic {3D} images,'' \emph{IEEE Transactions
  on Image Processing}, vol.~24, no.~11, pp. 3400--3414, 2015.

\bibitem{dayan1998hierarchical}
P.~Dayan, ``A hierarchical model of binocular rivalry,'' \emph{Neural
  Computation}, vol.~10, no.~5, pp. 1119--1135, 1998.

\bibitem{hohwy2008predictive}
J.~Hohwy, A.~Roepstorff, and K.~Friston, ``Predictive coding explains binocular
  rivalry: An epistemological review,'' \emph{Cognition}, vol. 108, no.~3, pp.
  687--701, 2008.

\bibitem{leopold1996activity}
D.~A. Leopold and N.~K. Logothetis, ``Activity changes in early visual cortex
  reflect monkeys' percepts during binocular rivalry,'' \emph{Nature}, vol.
  379, no. 6565, p. 549, 1996.

\bibitem{spratling2017review}
M.~W. Spratling, ``A review of predictive coding algorithms,'' \emph{Brain and
  cognition}, vol. 112, pp. 92--97, 2017.

\bibitem{friston2003learning}
K.~Friston, ``Learning and inference in the brain,'' \emph{Neural Networks},
  vol.~16, no.~9, pp. 1325--1352, 2003.

\bibitem{friston2005theory}
K.~Friston, ``A theory of cortical responses,'' \emph{Philosophical transactions of
  the Royal Society B: Biological sciences}, vol. 360, no. 1456, pp. 815--836,
  2005.

\bibitem{rao1999predictive}
R.~P. Rao and D.~H. Ballard, ``Predictive coding in the visual cortex: a
  functional interpretation of some extra-classical receptive-field effects,''
  \emph{Nature neuroscience}, vol.~2, no.~1, p.~79, 1999.

\bibitem{spratling2012predictive}
M.~W. Spratling, ``Predictive coding accounts for {V1} response properties
  recorded using reverse correlation,'' \emph{Biological Cybernetics}, vol.
  106, no.~1, pp. 37--49, 2012.

\bibitem{spratling2010predictive}
M.~W. Spratling, ``Predictive coding as a model of response properties in cortical area
  v1,'' \emph{Journal of neuroscience}, vol.~30, no.~9, pp. 3531--3543, 2010.

\bibitem{shipp2013reflections}
S.~Shipp, R.~A. Adams, and K.~J. Friston, ``Reflections on agranular
  architecture: predictive coding in the motor cortex,'' \emph{Trends in
  neurosciences}, vol.~36, no.~12, pp. 706--716, 2013.

\bibitem{srinivasan1982predictive}
M.~V. Srinivasan, S.~B. Laughlin, and A.~Dubs, ``Predictive coding: a fresh
  view of inhibition in the retina,'' \emph{Proceedings of the Royal Society of
  London. Series B. Biological Sciences}, vol. 216, no. 1205, pp. 427--459,
  1982.

\bibitem{kilner2007predictive}
J.~M. Kilner, K.~J. Friston, and C.~D. Frith, ``Predictive coding: an account
  of the mirror neuron system,'' \emph{Cognitive processing}, vol.~8, no.~3,
  pp. 159--166, 2007.

\bibitem{atal1979predictive}
B.~Atal and M.~Schroeder, ``Predictive coding of speech signals and subjective
  error criteria,'' \emph{IEEE Transactions on Acoustics, Speech, and Signal
  Processing}, vol.~27, no.~3, pp. 247--254, 1979.

\bibitem{vuust2009predictive}
P.~Vuust, L.~Ostergaard, K.~J. Pallesen, C.~Bailey, and A.~Roepstorff,
  ``Predictive coding of music--brain responses to rhythmic incongruity,''
  \emph{cortex}, vol.~45, no.~1, pp. 80--92, 2009.

\bibitem{sun2018large}
W.~Sun, K.~Gu, S.~Ma, W.~Zhu, N.~Liu, and G.~Zhai, ``A large-scale compressed
  360-degree spherical image database: From subjective quality evaluation to
  objective model comparison,'' in \emph{2018 IEEE 20th International Workshop
  on Multimedia Signal Processing (MMSP)}.\hskip 1em plus 0.5em minus
  0.4em\relax IEEE, 2018, pp. 1--6.

\bibitem{duan2018perceptual}
H.~Duan, G.~Zhai, X.~Min, Y.~Zhu, Y.~Fang, and X.~Yang, ``Perceptual quality
  assessment of omnidirectional images,'' in \emph{2018 IEEE International
  Symposium on Circuits and Systems (ISCAS)}.\hskip 1em plus 0.5em minus
  0.4em\relax IEEE, 2018, pp. 1--5.

\bibitem{yu2015framework}
M.~Yu, H.~Lakshman, and B.~Girod, ``A framework to evaluate omnidirectional
  video coding schemes,'' in \emph{2015 IEEE International Symposium on Mixed
  and Augmented Reality}.\hskip 1em plus 0.5em minus 0.4em\relax IEEE, 2015,
  pp. 31--36.

\bibitem{sun2017weighted}
Y.~Sun, A.~Lu, and L.~Yu, ``Weighted-to-spherically-uniform quality evaluation
  for omnidirectional video,'' \emph{IEEE signal processing letters}, vol.~24,
  no.~9, pp. 1408--1412, 2017.

\bibitem{zakharchenko2016quality}
V.~Zakharchenko, K.~P. Choi, and J.~H. Park, ``Quality metric for spherical
  panoramic video,'' in \emph{Optics and Photonics for Information Processing
  X}, vol. 9970.\hskip 1em plus 0.5em minus 0.4em\relax International Society
  for Optics and Photonics, 2016, p. 99700C.

\bibitem{kim2019deep}
H.~G. Kim, H.-t. Lim, and Y.~M. Ro, ``Deep virtual reality image quality
  assessment with human perception guider for omnidirectional image,''
  \emph{IEEE Transactions on Circuits and Systems for Video Technology}, 2019.

\bibitem{yang20183d}
J.~Yang, T.~Liu, B.~Jiang, H.~Song, and W.~Lu, ``{3D} panoramic virtual reality
  video quality assessment based on {3D} convolutional neural networks,''
  \emph{IEEE Access}, vol.~6, pp. 38\,669--38\,682, 2018.

\bibitem{ji20133d}
S.~Ji, W.~Xu, M.~Yang, and K.~Yu, ``{3D} convolutional neural networks for
  human action recognition,'' \emph{IEEE transactions on pattern analysis and
  machine intelligence}, vol.~35, no.~1, pp. 221--231, 2013.

\bibitem{xu2018assessing}
M.~Xu, C.~Li, Z.~Chen, Z.~Wang, and Z.~Guan, ``Assessing visual quality of
  omnidirectional videos,'' \emph{IEEE Transactions on Circuits and Systems for
  Video Technology}, 2018.

\bibitem{xu2018predicting}
M.~Xu, Y.~Song, J.~Wang, M.~Qiao, L.~Huo, and Z.~Wang, ``Predicting head
  movement in panoramic video: A deep reinforcement learning approach,''
  \emph{IEEE transactions on pattern analysis and machine intelligence}, 2018.

\bibitem{yang2017objective}
S.~Yang, J.~Zhao, T.~Jiang, J.~W.~T. Rahim, B.~Zhang, Z.~Xu, and Z.~Fei, ``An
  objective assessment method based on multi-level factors for panoramic
  videos,'' in \emph{2017 IEEE Visual Communications and Image Processing
  (VCIP)}.\hskip 1em plus 0.5em minus 0.4em\relax IEEE, 2017, pp. 1--4.

\bibitem{rai2017saliency}
Y.~Rai, P.~Le~Callet, and P.~Guillotel, ``Which saliency weighting for omni
  directional image quality assessment?'' in \emph{2017 Ninth International
  Conference on Quality of Multimedia Experience (QoMEX)}.\hskip 1em plus 0.5em
  minus 0.4em\relax IEEE, 2017, pp. 1--6.

\bibitem{rai2017dataset}
Y.~Rai, J.~Guti{\'e}rrez, and P.~Le~Callet, ``A dataset of head and eye
  movements for 360 degree images,'' in \emph{Proceedings of the 8th ACM on
  Multimedia Systems Conference}.\hskip 1em plus 0.5em minus 0.4em\relax ACM,
  2017, pp. 205--210.

\bibitem{xu2018subjective}
J.~Xu, C.~Lin, W.~Zhou, and Z.~Chen, ``Subjective quality assessment of
  stereoscopic omnidirectional image,'' in \emph{Pacific Rim Conference on
  Multimedia}.\hskip 1em plus 0.5em minus 0.4em\relax Springer, 2018, pp.
  589--599.

\bibitem{vaseghi2008advanced}
S.~V. Vaseghi, \emph{Advanced digital signal processing and noise
  reduction}.\hskip 1em plus 0.5em minus 0.4em\relax John Wiley \& Sons, 2008.

\bibitem{friston2002functional}
K.~Friston, ``Functional integration and inference in the brain,''
  \emph{Progress in neurobiology}, vol.~68, no.~2, pp. 113--143, 2002.

\bibitem{kersten2004object}
D.~Kersten, P.~Mamassian, and A.~Yuille, ``Object perception as bayesian
  inference,'' \emph{Annu. Rev. Psychol.}, vol.~55, pp. 271--304, 2004.

\bibitem{sheikh2006statistical}
H.~R. Sheikh, M.~F. Sabir, and A.~C. Bovik, ``A statistical evaluation of
  recent full reference image quality assessment algorithms,'' \emph{IEEE
  Transactions on image processing}, vol.~15, no.~11, pp. 3440--3451, 2006.

\bibitem{zhu2016stereoscopic}
Y.~Zhu, G.~Zhai, K.~Gu, Z.~Che, and D.~Li, ``Stereoscopic image quality
  assessment with the dual-weight model,'' in \emph{2016 IEEE International
  Symposium on Broadband Multimedia Systems and Broadcasting (BMSB)}.\hskip 1em
  plus 0.5em minus 0.4em\relax IEEE, 2016, pp. 1--6.

\bibitem{abbas2016gopro}
A.~Abbas, ``{GoPro} test sequences for virtual reality video coding,'' in
  \emph{Doc. JVET-C0021, Joint Video Exploration Team (on Future Video coding)
  of ITU-T VCEG and ISO/IEC MPEG, Geneva, CH, 3rd meeting}, 2016.

\bibitem{itu1999subjective}
P.~ITU-T~RECOMMENDATION, ``Subjective video quality assessment methods for
  multimedia applications,'' \emph{International telecommunication union},
  1999.

\bibitem{moorthy2013subjective}
A.~K. Moorthy, C.-C. Su, A.~Mittal, and A.~C. Bovik, ``Subjective evaluation of
  stereoscopic image quality,'' \emph{Signal Processing: Image Communication},
  vol.~28, no.~8, pp. 870--883, 2013.

\bibitem{brunnstrom2009vqeg}
K.~Brunnstrom, D.~Hands, F.~Speranza, and A.~Webster, ``{VQEG} validation and
  {ITU} standardization of objective perceptual video quality metrics
  [standards in a nutshell],'' \emph{IEEE Signal processing magazine}, vol.~26,
  no.~3, pp. 96--101, 2009.

\bibitem{zhang2014vsi}
L.~Zhang, Y.~Shen, and H.~Li, ``{VSI}: A visual saliency-induced index for
  perceptual image quality assessment,'' \emph{IEEE Transactions on Image
  Processing}, vol.~23, no.~10, pp. 4270--4281, 2014.

\bibitem{he2016jvet}
Y.~He, X.~Xiu, Y.~Ye, V.~Zakharchenko, and E.~Alshina, ``{JVET 360Lib} software
  manual,'' \emph{Joint Video Exploration Team (JVET) of ITU-T SG16 WP3 and
  ISO/IEC JTC1/SC29/WG11}, 2016.

\bibitem{hewage2010reduced}
C.~T. Hewage and M.~G. Martini, ``Reduced-reference quality metric for {3D}
  depth map transmission,'' in \emph{2010 3DTV-Conference: The True
  Vision-Capture, Transmission and Display of 3D Video}.\hskip 1em plus 0.5em
  minus 0.4em\relax IEEE, 2010, pp. 1--4.

\bibitem{bensalma2013perceptual}
R.~Bensalma and M.-C. Larabi, ``A perceptual metric for stereoscopic image
  quality assessment based on the binocular energy,'' \emph{Multidimensional
  Systems and Signal Processing}, vol.~24, no.~2, pp. 281--316, 2013.

\end{thebibliography}

%

%
%
%




\end{document}